\documentclass[12pt, a4paper]{article}
\pdfoutput=1
\usepackage{jcappub} 

\usepackage{datetime}
\usepackage{amsmath}
\usepackage{amsfonts}
\usepackage{amssymb}
\usepackage{latexsym}
\usepackage{graphicx}
\usepackage{grffile}
\usepackage{color}
\usepackage{mathrsfs}
\usepackage{slashed}
\usepackage{hyperref}
\usepackage{enumerate}
\usepackage{multirow}
\usepackage[table]{xcolor}
\usepackage{colortbl}
\usepackage{url}
\usepackage{enumitem}
\usepackage{comment}
\usepackage{bm}
\usepackage{caption}
\usepackage{subcaption}
\usepackage{float}

\numberwithin{equation}{section}

\definecolor{rossos}{rgb}{0.7,0,0.3}
\definecolor{violachiaro}{rgb}{1,0.6,1}
\definecolor{rossochiaro}{rgb}{1,0.6,0.6}
\definecolor{verdechiaro}{rgb}{0.6,1,0.6}
\definecolor{giallochiaro}{rgb}{1,1,0.6}
\definecolor{bluscuro}{rgb}{0.15, 0.2, 0.9}
\definecolor{verdes}{rgb}{0.1, 0.5, 0.1}
\definecolor{gold}{rgb}{1,0.84,0}
\definecolor{forestgreen}{rgb}{0.13,0.55,0.13}
\def\hhref#1{\href{http://arxiv.org/abs/#1}{#1}} 

\makeatother   

\definecolor{oucrimsonred}{rgb}{0.6, 0.0, 0.0}
\definecolor{persianblue}{rgb}{0.11, 0.22, 0.73}
\definecolor{forestgreen}{rgb}{0.13,0.35,0.13}
 \hypersetup{colorlinks, citecolor=oucrimsonred, linkcolor=persianblue, urlcolor=oucrimsonred}

 \def\be   {\begin{equation}}   \def\ee   {\end{equation}}
 \def\ba   {\begin{array}}      \def\ea   {\end{array}}
 \def\bea  {\begin{eqnarray}}   \def\eea  {\end{eqnarray}}
 \def\bean {\begin{eqnarray*}}  \def\eean {\end{eqnarray*}}

\begin{document}
\begin{flushright}
{\footnotesize
{\sc SACLAY-T15/165}
}
\end{flushright}

\title{Towards a realistic astrophysical interpretation of the {gamma-ray} Galactic center excess.}

\author[a,b]{Daniele Gaggero}
\affiliation[a]{SISSA, via Bonomea 265, 34136 Trieste, Italy}
\affiliation[b]{INFN, Sezione di Trieste, via Valerio 2, 34127 Trieste, Italy}

\author[c]{Marco Taoso}
\affiliation[c]{{Institut de Physique Th\'eorique}, Universit\'e Paris Saclay, CNRS, CEA, F-91191 Gif-sur-Yvette, France}

\author[a,b]{Alfredo Urbano}

\author[a,b]{Mauro Valli}

\author[a,b]{Piero Ullio}

\emailAdd{daniele.gaggero@sissa.it}
\emailAdd{marco.taoso@cea.fr}
\emailAdd{alfredo.urbano@sissa.it}
\emailAdd{mauro.valli@sissa.it}
\emailAdd{piero.ullio@sissa.it}


\date{\today}

\abstract{ 


A spherical-symmetric gamma-ray emission from {the inner region of the Galaxy (at least up to roughly $10^\circ$ in latitude and longitude)} has been recently identified in Fermi-LAT data, and initially associated to dark matter particle annihilations. 
Guided by the evidence for a high gas density in the inner kpc of the Galaxy correlated with a very large Supernova rate, and hence with ongoing cosmic-ray acceleration, we investigate instead the possibility of addressing this excess in terms of ordinary cosmic-ray sources and standard steady-state diffusion. 
We {alter the source term, and consistently the correlated gamma-ray emissions, in the context of a template-fitting analysis. We focus on a region of interest (ROI) defined as:
$|l|<20^{\circ}; \hspace{5mm} 2^{\circ}<|b|<20^{\circ},$ with $l$ and $b$ the Galactic longitude and latitude coordinates.}
We analyze in detail the overall goodness of the fit of our framework, and perform a detailed direct comparison against data examining profiles in different directions. 
{Remarkably, the test statistic of the fit related to our scenario turns out to be as good as the Dark Matter one in the ROI here considered.} 



}


\maketitle
\flushbottom


\section{Introduction}

There is a remarkable agreement between models for the diffuse $\gamma$-ray emission in the Galaxy and data from the all-sky survey from the Fermi Large Area Telescope. A fairly good match~\cite{galprop2012} is obtained in most region of the sky implementing only minor readjustments to a standard recipe based on: {\sl i)} supernova remnants (SNRs) as cosmic-ray (CR) sources; {\sl ii)} the steady-state propagation of CRs in the Galaxy as tuned on local CR measurements; and {\sl iii)} $\gamma$-ray emitting targets, namely the gas and the interstellar radiation field (ISRF), as indirectly derived from observations at other wavelengths. 

Notable exceptions are: A systematic underestimate of the flux above the GeV in the inner Galactic plane~\cite{galprop2012,RadialGradients}; A hard-spectrum emission extending about $40$ degrees in longitude and $50$ degrees above and below the Galactic center (GC) commonly dubbed Fermi bubbles~\cite{fermibubbles} -- most likely associated to the the signal with similar morphology previously discovered at microwave frequencies and dubbed WMAP haze~\cite{wmaphaze}; A discrepancy between the predicted CR gradient and the one inferred from $\gamma$-ray emissivity along the Galactic plane \cite{Evoli:2012ha};  A roughly spherical excess in the inner few degrees from the GC and peaking at an energy of few GeV~\cite{excess_history}.

The GC region, namely the inmost few hundred pc, is one of the most challenging from the point of view the theoretical modelling for the diffuse $\gamma$-ray emissivity. All the three ingredients we listed above for the standard recipe show problematic aspects. First of all, catalogues used for CR source distributions, mainly the observed distributions of SNRs~\cite{casebhattacharya} and pulsars~\cite{lorimer,yusinof}, are not optimal for the inner Galaxy. The fitting functions extrapolated from such catalogues, and used for CR transport studies and to derive $\gamma$-ray emissivities in numerical propagation codes such as {\tt Galprop}~\cite{GalpropWeb,Vladimirov2012,galprop2000,galprop2002} and {\tt DRAGON}~\cite{DragonWeb,evoli:2008,gaggero:2013}, simply go to zero at the GC and fail to account for the very active star forming region emerging from multi-wavelength surveys of the central Galaxy. 
From the point of view of CR propagation, there are several features making it likely that transport properties in the GC region differ significantly from average properties in the Galaxy as fitted to local CR data. E.g., the GC region seems to host strong magnetic fields, at the level of $50$-$100$~$\mu$G~\cite{crockernature}, much above the estimated average value within the disc or the CR diffusive halo, probably impacting on the structure and effective values for the diffusion tensor. Also, very strong convective effects have been claimed in this region~\cite{crocker2011a,crocker2011b}, in analogy to outflows from external starburst systems, as well as from the fact that at the GC there is a violation of the correlation between far-infrared and radio continuum luminosities observed for systems with in-situ energy losses of CR electrons. Finally, the density distribution and dynamics of molecular gas at the GC is still actively investigated, see, e.g.,~\cite{ott2014}.

The evidence for a GC excess has been advocated based on the template-fitting method. Extensively used for the analysis and subtraction of CMBR foregrounds, this technique has led first to the identification of the WMAP haze and then successfully applied to search for its (alleged) $\gamma$-ray counterpart, the Fermi bubbles. The rationale of this procedure is to test whether, on top of small-scale discrepancies, the model for diffuse $\gamma$-ray emission intercepts the correct intermediate- to large-scale morphologies, or data favour extra contributions with different angular imprint in the region under investigation. Morphological templates are assigned -- based on theoretical modelling and/or observations -- to components connected to different physical emission mechanisms, such as: The component taken as the sum of the term due to the production and decay of $\pi^0$ plus bremsstrahlung emission, both related to the gas distribution in the Galaxy;  the Inverse Compton (IC) term, as correlated to the ISRF model; and the isotropic template for the extra-Galactic background component. Each template is then let free to fluctuate, independently for each energy bin, to find the configuration providing the best fit to the data; in this way one accounts for possible spectral distortions and normalisation uncertainties of the different components within the theoretical model. Indications for an extra component can be claimed if a significant improvement of the overall fit is found when repeating the same exercise with an additional physically motivated template.

In case of the GC the application of the template fitting procedure has led to the identification of residuals with an approximately spherical morphology; such excess can be mitigated, and at least partially reabsorbed, introducing in the analysis an extra component which has indeed a strong physical motivation: annihilating dark matter (DM) particles. In fact, the Milky Way is believed to be supported by an extended halo of DM particles, whose density should be maximal in the GC region; if such particles can annihilate in pairs, like in many models for particle-physics motivated DM candidates, they would generate a $\gamma$-ray flux peaking at the GC. The template morphology, as well as the spectral shape found as a result of the template fitting procedure, have been claimed to be compatible with such a DM signal, pointing to the very appealing scenario of Weakly Interacting Massive Particles (WIMPs) as early Universe thermal relics~\cite{excess_history,Daylan:2014rsa}. Several theoretical embeddings of this picture from the DM model-building side were proposed~\cite{excess model building}, as well as its consistency with results from other WIMP searches has been tested~\cite{excess constraints}.

On the other hand, a few competing scenarios accounting for the extra component have been suggested as well. Among the proposed explanations there are: the presence a population of unresolved millisecond pulsars~\cite{MSP,Petrovic:2014uda}, a bursting star-forming activity in the past near the GC~\cite{Carlson:2014cwa}, or nonthermal bremsstrahlung produced by a population of electrons interacting with neutral gas in molecular clouds~\cite{YusefZadeh:2012nh,Macias:2013vya}. At the moment is still unclear whether there is a preferred scenario and further investigations are ongoing. E.g., concerning the MSP explanation, while there seems to be compatibility with the average spectrum of observed MSPs, the consistency with the number of pulsars observed by Fermi-LAT is still debated. Interestingly, two recent analyses aiming at an improved targeting of point sources~\cite{Bartels:2015aea,Lee:2015fea} suggest that the GC excess could be produced by a population of unresolved point sources. Also very recently, the burst scenario was considered in great detail in \cite{Cholis:2015dea}; the conclusion of this work is that a combination of two burst events may provide a good fit of the excess spectrum, although with quite unnatural values of the involved parameters (e.g. extremely hard spectral indices) and without reproducing the observed morphology of the excess emission from the innermost degree or two around the GC.

In this analysis we concentrate instead on the choice of templates for the CR-induced components and on the impact of such choice on the evidence for the excess. As discussed above, none of the ingredients of the ``standard'' theoretical model setting those templates are actually optimised for the GC environment. As a preliminary step we tested a few hypotheses on whether significant changes in the CR transport equation in GC region, including a sampling of radial dependences for the diffusion coefficient and anisotropic diffusion, would mimic the required morphological features; at this level, in agreement with and extension of results in~\cite{Zhou:2014lva,Calore}, we did not find suitable configurations. 

We have turned then to considering an adjustment in the choice of the CR source function, introducing, as a new ingredient, a steady-state source located at the GC and with a narrow spatial extension -- few hundred pc. As already mentioned above, there are robust astrophysical observations supporting such new ingredient: The GC seems to harbour significant star formation and a large rate of Supernova explosions compared to the average value in the Galaxy; according to \cite{star_formation_GC}, the star formation rate (SFR) in the inner few degrees away from the GC is of order $1$\%  of the SFR in the Galaxy, making it roughly a factor of $250$ higher than the mean rate in the Galaxy. This should be the consequence of the following two facts: The presence a large reservoir of molecular gas filling the inner part of the Galactic bulge (dubbed Central Molecular Zone~\cite{central_molecular_zone} and extending up to $\sim 200\div300$ pc away from the GC); the very peculiar physical properties of this environment, where the insterstellar medium appears significantly hotter, more turbulent, and more magnetized. Solid evidence for the high SFR level comes from infrared observations -- performed with the Hubble Space Telescope -- of some extremely dense stellar clusters in the inner 50 pc (the Central, Arches, and Quintuplet clusters). These structures are rich of young, very hot stars that are many times larger and more massive than the Sun, see, e.g., \cite{Figer:2002bi,Figer:2008kf} and references therein. Moreover, many isolated Wolf-Rayet Stars and O Supergiants were observed in the inner 100 pc~\cite{Mauerhan:2010ih}. The Supernova explosions connected with this relevant star formation activity are expected to accelerate a large amount of CRs. While this contribution has no impact on the local CR fluxes (see, e.g., the discussion on the locality of CR spectra in \cite{Evoli:2011id}), it drives major consequences on $\gamma$-ray emissivities in the GC region, as we will discuss in the rest of the paper.

At the level of proper statistical assessments of the goodness of the fit to the data, a comparison between alternative theoretical models is a very hard task due to small-scale discrepancies which have a significant impact on statistical estimators and prevent a reliable comparison with the data. In order to alleviate these problems, we have adopted therefore the template-fitting strategy. 
We take as reference case the fit of the data for ``standard-lore'' CR templates, rederive the improvements in the fit obtained when including the DM template, and then compare such refinements to those we can obtain in the case when the new CR templates are adopted (different methods on how to weight fit improvements will be discussed). From such a comparison we will show that the two scenarios are in most respects equivalent and hence that the ``excess'' can be reabsorbed within a framework involving only a minimal picture of steady-state propagation of CRs. 
{{Note that the approach we follow is the appropriate one to realistically reproduce the energy dependence of the diffuse emission template: Indeed, while the Dark Matter morphology is fixed at all energies, the IC, $\pi_{0}$ and bremsstrahlung ones change in energy 
as a result of CR transport. It is therefore extremely important to incorporate the contribution related to the new source directly into the original diffuse emission template.}}





\section{Description of the method}
\label{sec:method}

\subsection{The template-fitting procedure}

In this section we describe 
the method that we use to compare our proposed astrophysical scenario with the DM interpretation of the GC excess.

The starting point of the analysis is a physical model for the CR distribution in the Galaxy obtained with {\tt DRAGON}~\cite{evoli:2008,gaggero:2013}. This numerical code is designed to simulate all processes related to Galactic CR transport (in particular: diffusion, reacceleration, convection, energy losses, spallation) for all CR species, from heavy nuclei to protons, antiprotons, and leptons.
{\tt DRAGON} can work in both 2D and 3D mode, and in both cases it is possible to implement anisotropic diffusion. 
The code includes the nuclear cross section database and interstellar radiation field model taken from the latest public version of {\tt Galprop}~\cite{GalpropWeb}.

We then perform a line-of-sight integral and obtain the $\gamma$-ray skymaps from $\simeq 0.3$ GeV to $\simeq 300$ GeV using {\tt GammaSky}~\cite{GammaSky}. 
This package computes the $\gamma$-ray emission due to: {\it i)} decay of neutral pions produced by collision of CRs with the interstellar gas; {\it ii)} IC scattering of CR leptons on the diffuse radiation field; {\it iii)} Bremsstrahlung emission due to CR leptons interacting with the interstellar gas.
%
Exploiting such description 
of the diffusion emission induced by the CR interactions, we model the $\gamma$-ray sky as a linear superposition of the following templates:

\begin{itemize}
\item $\pi^0$-decay $+$ Bremsstrahlung template (produced with {\tt GammaSky}). 
These templates are generated making use of the CR distributions computed with {\tt DRAGON} and adopting the gas model taken from the public version of the {\tt Galprop} CR package~\cite{GalpropWeb,Vladimirov2012,galprop2000,galprop2002}.

\item IC template (produced with {\tt GammaSky}). Here we use the leptonic distributions calculated with {\tt DRAGON} and the most recent ISRF model included in {\tt Galprop}. 

\item Isotropic extragalactic background template (EGB) with a spectrum taken from a recent analysis of the Fermi collaboration~\cite{Ackermann:2014usa}.

\item Point source (PS) template. This ingredient is obtained from the 4-year Point Source Catalog (3FGL) provided by the Fermi collaboration~\cite{TheFermi-LAT:2015hja}. The angular resolution of the Fermi-LAT instrument is taken into account smoothing the emission of each point source according to the point spread function (PSF) of the instrument. The PSF is modeled using the Fermi tool {\tt gtpsf}.

\item {\it Fermi bubbles} template with morphology and spectrum as derived by the Fermi collaboration in~\cite{Fermi-LAT:2014sfa}.

\item A DM template  obtained integrating along the line of sight the square of a generalized NFW profile: $$\rho(r)=\rho_0 \left(\frac{r}{r_s} \right)^{-\gamma} \left(1+\frac{r}{r_s}\right)^{-3+\gamma},$$
where $r$ corresponds is the galactocentric distance and the scale radius $r_s$ is fixed to $r_s=20$ kpc. The normalization $\rho_0$ is chosen in order to obtain a DM density at the Sun position $\rho(r_{\odot})=0.3$ GeV cm$^{-3}$, with $r_{\odot}=8.5$ kpc.
For the inner slope we take $\gamma = 1.26,$ as in~\cite{Daylan:2014rsa}.
We checked that different values of $\gamma$ do not alter the conclusion of our analysis. In general, we found that profiles with $\gamma \simeq 1-1.6$ are preferred by the data.

\end{itemize}

We consider $5$ years of Fermi-LAT data, within the event class {\tt ULTRACLEAN}. We use the Fermi tools {\tt v9r32p5} to perform analysis cuts and to compute the exposure of the instrument.
Events with zenith angles larger than 100$^\circ$ are rejected. We also apply the following  selection cuts:
{\tt (DATA\textunderscore QUAL==1) \&\& (LAT\textunderscore CONFIG==1) \&\& (ABS(ROCK\textunderscore ANGLE)<52) \&\& (IN\textunderscore SAA!=T)}. 
The exposure maps are computed using the response function {\tt P7REP\textunderscore ULTRACLEAN\textunderscore V15}.
The data are organized in $30$ energy bins from $300$ MeV to $300$ GeV, equally spaced on a logarithmic scale.

\medskip

We work with the {\tt HEALPix} tessellation scheme for all our maps~\cite{Gorski:2004by}. We choose a resolution $n_{side}=256,$ corresponding to a pixel size of $\sim 0.23^{\circ}.$
Finally, the templates and the Fermi-LAT data are smoothed to a common angular size using a gaussian kernel. In this way all the maps have the same angular resolution. For this purpose we use the {\tt HEALPix} routine {\tt smoothing}. We follow the method in~\cite{WeiEAlfredo} using $f_{target}=3^{\circ}$ for $E_{\gamma}<0.6$ GeV and $f_{target}=1^{\circ}$ for $E_{\gamma}>0.6$. 
We checked that our results are insensitive to such smoothing choices, {with the exception of the low-energy portion of the spectrum where the recipe adopted for the smoothing of the templates has a non negligible impact on the result, see below for further details.}

\medskip

At the end of this process, we perform the template-fitting analysis. We focus on a region of interest (ROI) defined as:
$$|l|<20^{\circ}; \hspace{5mm} 2^{\circ}<|b|<20^{\circ},$$ with $l$ and $b$ the longitude and latitude coordinates.
For each energy bin, we construct then the following Poisson likelihood function:
\begin{equation}
-2 \ln(\mathcal{L}) \, = \,  \, 2 \sum_{i} {  \left( e_{i} - o_{i} ~\ln(e_{i}) \right) \,+\,2 \sum_i \ln(o_{i}!) \,+\, \chi^2_{\rm ext}}\, ,
\label{eq:Likelihood}
\end{equation}
where the index $i$ labels each pixel inside the ROI, and we introduce the following quantities:
\begin{itemize}

\item $o_i$ is the number of {\it observed} events (the Fermi-LAT data);

\item $e_i$ is the number of {\it expected} events. This term is written as a linear combination of the events corresponding to the templates described above

\begin{equation}
e_i =~\sum_k( \alpha^k e_i^k)
\label{eq:alphak}
\end{equation}

where $k$ runs over the different templates. The counts are obtained multiplying the differential flux per unit of solid angle by the Fermi-LAT exposure map, the energy bin width and the solid angle per pixel.

\item The Fermi bubbles and the isotropic emissions are better determined with observations outside our ROI.
For this reason, we introduce a penalty factor $\chi^2_{\rm ext}$ for their coefficients:
\begin{equation}
\chi^2_{\rm ext}= \sum_{h} \frac{\phi_h^2 \left(1 -\alpha^h \right)^2}{\epsilon_h^2}.
\end{equation}
where $h$ now only refers to the Fermi bubbles and isotropic emission templates. $\phi_h$ and $\epsilon_h$ are the nominal fluxes and uncertainties taken from the Fermi analysis mentioned above.

\end{itemize}
The minimization of the function \ref{eq:Likelihood} provides the best-fit values of the coefficients $\alpha^k$ in each energy bin. 

%

Concerning the treatment of point-source emission, we checked the possibility of introducing a masking procedure to mitigate their role in our analysis and/or account for reconstruction faults. Masks of two types have considered: A sharp cut around each point source of all pixel within a 68$\%$ containment radius, depending on energy and sharply increasing at $E\lesssim1$~GeV, as obtained from the PSF of the instrument Fermi tool {\tt gtpsf}. We also considered the soft mask introduced in~\cite{Calore}, according to which the contribution of each pixel to the likelihood is reweigthed depending on the flux from the point sources expected in that pixel, lowering the impact on the likelihood of pixels whose counts are dominated by point sources. We found, however, that our results are not severely affected by adopting any of the two masks and hence disregard the masking procedure for simplicity.

\subsection{A novel reference model}

Having set the framework, we introduce a benchmark case to examine whether, within the template-fitting analysis, a properly modified CR emissivity model can be considered equivalent to introducing the DM template. 

We choose as starting point for the CR propagation framework one of the reference cases in \cite{Calore}, labelled ``Model A'' in that analysis. 

This model has a standard scaling of the diffusion coefficient with rigidity (a Kolmogorov-type $\delta = 0.33$), and reasonable values for the normalization of the diffusion coefficient ($5\cdot10^{28}\,{\rm cm^2/s}$ at $4$ GV), the height of the diffusion halo ($4$ kpc), and the Alfv\'en speed ($32.7$ km/s); moreover, the model features a moderate convective wind gradient ($50$ km/s/kpc).
{The injection slope is a broken power law. For the hadrons the indexes are $1.89 \, (2.47)$ below (above) the reference rigidity of $11.3$ GV, while for the electrons $1.6 \, (2.43)$ below (above) $2.18$ GV.} 

On the other hand, the model has a sharply enhanced IC component obtained by rescaling the normalisation of the CR electrons by a factor $\simeq 2$ with respect to local measurements, as well as a $\simeq 40$\% increase in the ISRF; the synchrotron energy losses are also enhanced, adjusting the overall normalisation of a conventional large-scale magnetic field model for the Galaxy in such way that, after the rescaling, it matches the claimed value of about 50~$\mu$G at the GC. While the latter assumptions make the model fail to be compatible with local CR measurements, we remind that the goal of the template-fitting analysis is not to find a model valid in the whole Galaxy, but rather to describe in detail the morphology of the GC region (reproducing also the local measurements, and possibily additional observables, would require instead a completely different method of analysis, clearly well beyond our scope; such an improvement will be considered in a forthcoming work).

As also discussed in  \cite{Calore}, the specific choice of propagation parameters is not in any way crucial for the template-fitting analysis. The point is that the free floating of the individual templates in each energy bin always tends to readjust the input model towards a preferred output, carrying in the end common features, such as an enhancement in the IC component.  Model A seems to be preferred in \cite{Calore} since implies the smallest changes in normalisation and spectral features between input and output model (namely the minimum of the likelihood function is reached for the smallest departure from $1$ of the coefficients $\alpha^k$ introduced in Eq.~\ref{eq:alphak}). It is also one of the cases in which the template fitting gives a very clean indication, over a large span of energy bins, for a sharp improvement of the data fit when including the DM template; we take the challenge to focus on this sort of ``worst-case scenario" to test whether our alternative picture can perform equally well. On the other hand, we stress again that, as tested on a few sample cases, changing the propagation setup would only impact on minor details of the fit, leaving the overall picture unchanged. 

Model A adopts as CR source function a smooth interpolation from a SNR catalogue \cite{casebhattacharya}. As already mentioned in the Introduction, this term clearly does not include a satisfactory description of the Galactic bulge region. We consider then an extra steady-state source term, which we model as a Gaussian (hereafter ``spike''):
\begin{equation}
  Q_{\rm spike} \, = \, Q_0 \exp{\left( -\frac{r^2}{\sigma^2} \right) }\,.
  \label{eq:spike}
\end{equation}

In the following, we will express the normalisation of this source in terms of $\mathcal{N}$, i.e. the ratio (in percentage) between the volume integral of the spike and the volume integral of the conventional source term. This provides a simple way to compare against observational constraints on the CR injection term in the center, correlated to the Star formation rate, or the Supernova explosion rate. 

Therefore, the most relevant parameters describing the new ingredient under consideration are then the spatial extension $\sigma$ and the normalisation $\mathcal{N}$ of the spike.

\begin{figure}[!htb!]
\begin{center}
\minipage{0.5\textwidth}
 \includegraphics[scale=0.4]{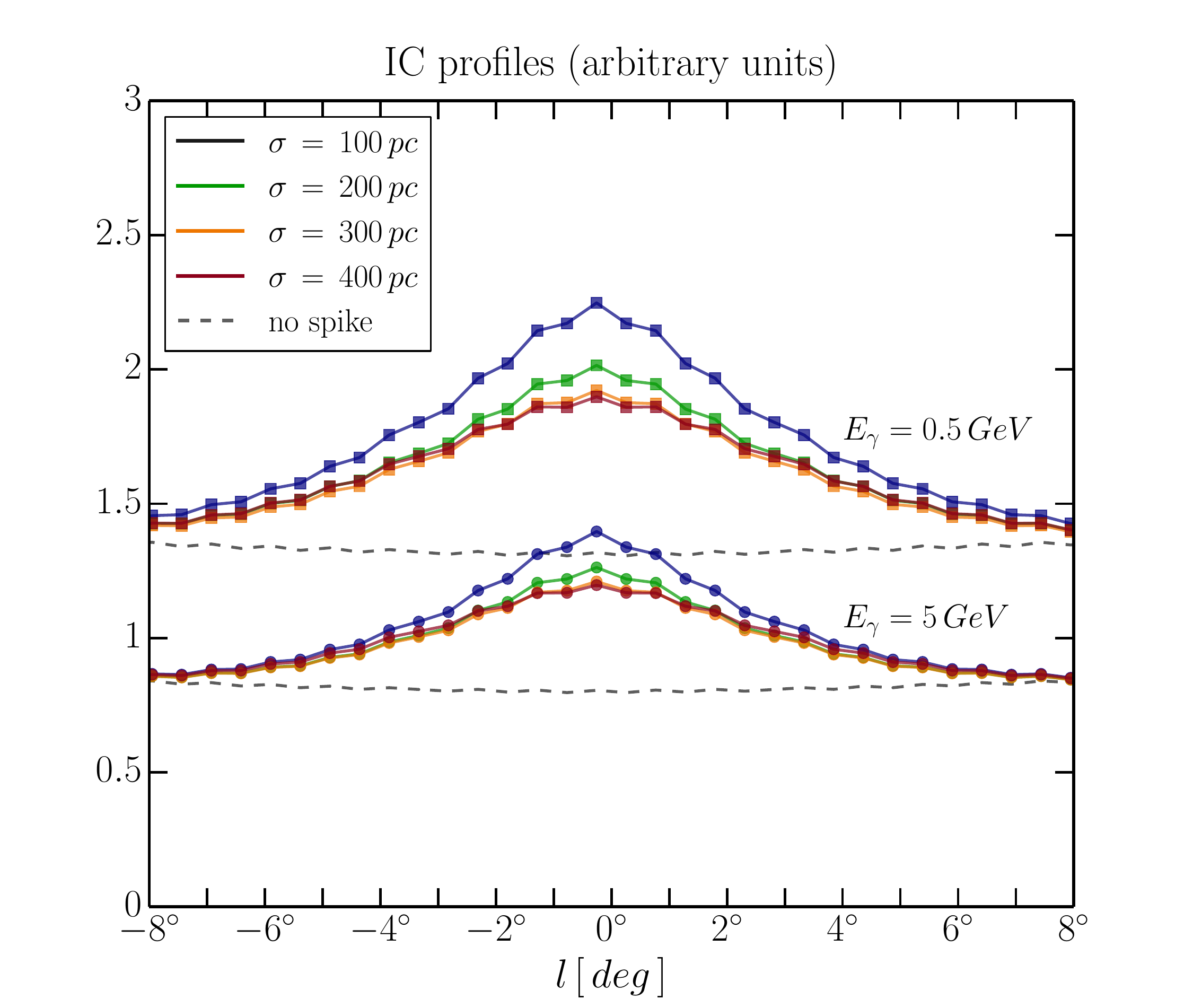}
\endminipage\hfill
\minipage{0.5\textwidth}
 \includegraphics[scale=0.4]{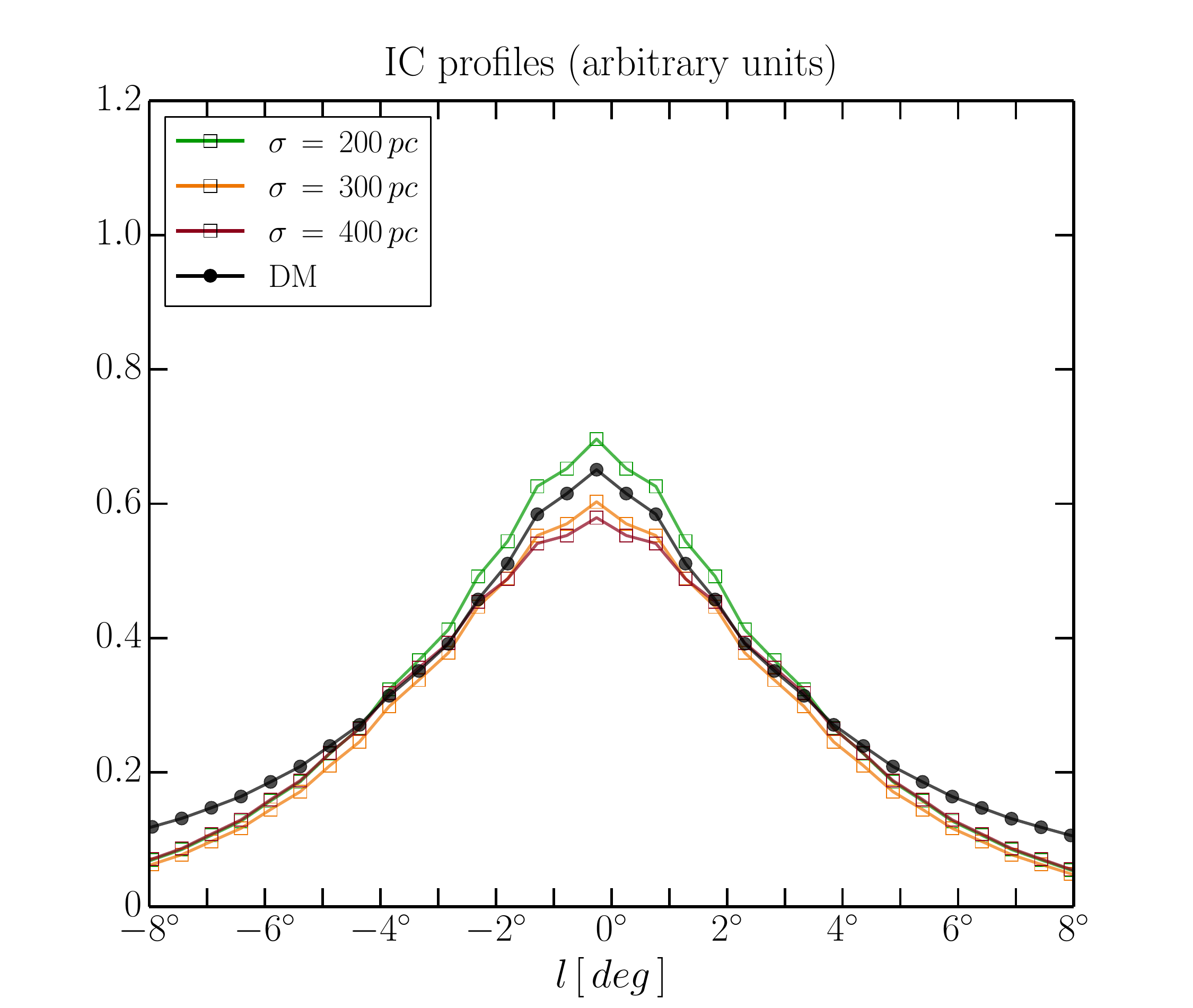}
\endminipage
\caption{\textit{{\bf Left panel.} The IC longitudinal profiles integrated in the range $2^\circ < |b| < 5^\circ$ for two reference values
of $E_{\gamma}$ and for spikes with different widths ($\sigma = 200 ~\div ~400 \,{\rm pc}$).
We remind that the normalization of the spike is set in such a way to absorb most of the GC excess. 
We checked {\it a posteriori} that the chosen value is compatible with astrophysical estimates regarding the star formation rate in the galactic center region. {\bf Right panel.} The IC profiles at $0.5$ GeV for the spike terms with different values of $\sigma$ compared to the DM template.}}\label{fig:IC_profile}
\end{center}
\end{figure}

In our ROI the spike mainly affects the IC template, since the region containing most of the $\pi^0$ and bremsstrahlung emission is masked out~\footnote{
{We remark that the reference model here considered is not tuned for the Galactic plane, within $2^\circ$ in latitude. In this region, a much more detailed treatment of the astrophysical ingredients (concerning both the emitting targets and the CR transport properties) would be in order; however, it is beyond the scope of our work.}} 
The additional contribution due to the spike impacts on the IC template as shown in the left panel of fig.~\ref{fig:IC_profile} where we plot the total IC longitudinal profile for different spike widths. On the right panel, we show instead the contribution to the IC emission from the spike only and compare it to the DM template. The plots are produced for sample $\gamma$-ray energy values.

It is important to notice that, while on the morphological side the two scenarios are similar, in the spike case we are not treating the extra ingredient as an independent term, rather we are correlating its spectrum to the overall IC emission.



\section{Results}
\label{sec:ReferenceCase}

\begin{figure}[!htb!]
\minipage{0.5\textwidth}
  \includegraphics[width=1. \linewidth]{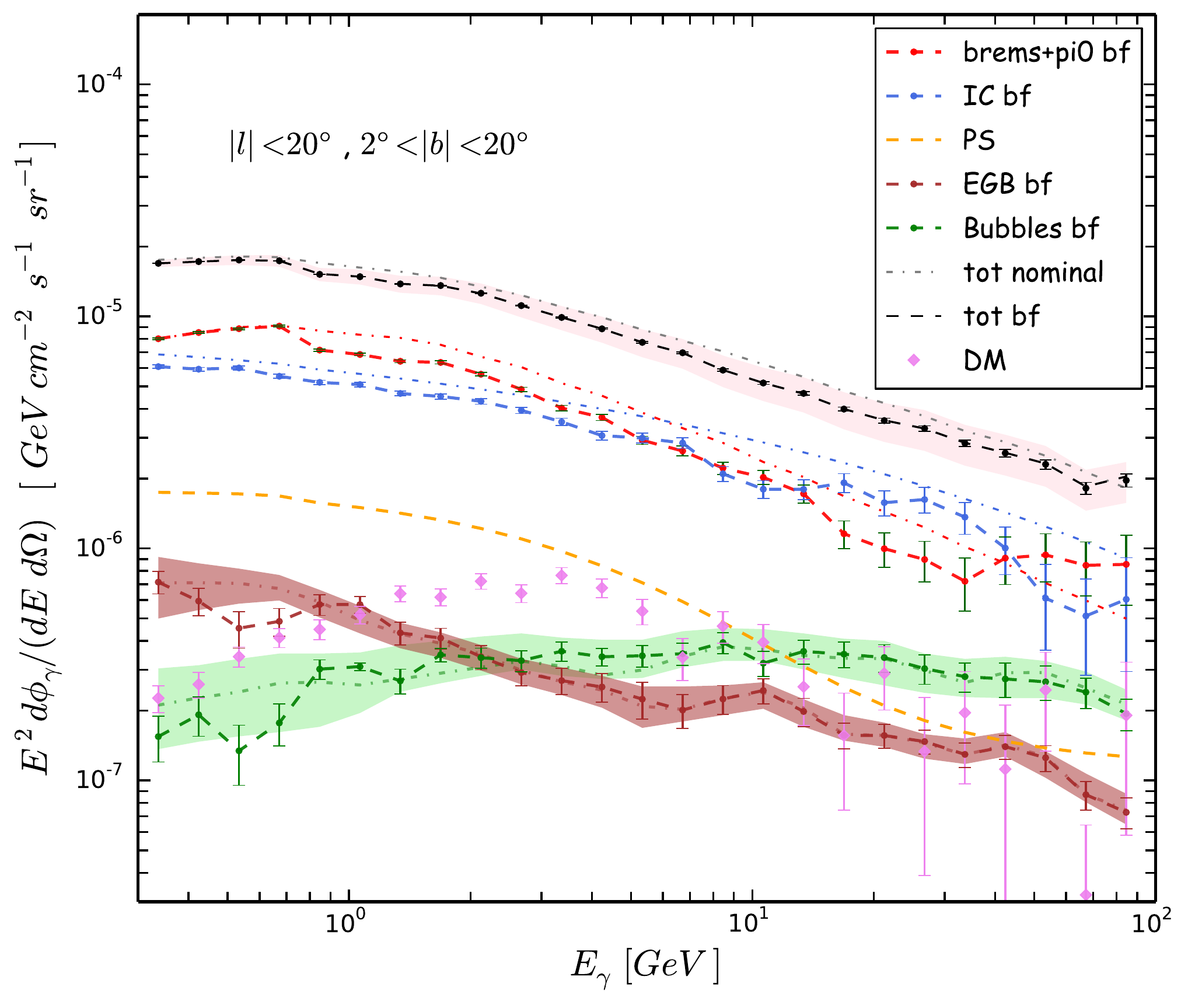}
\endminipage\hfill
\minipage{0.5\textwidth}
  \includegraphics[width=1. \linewidth]{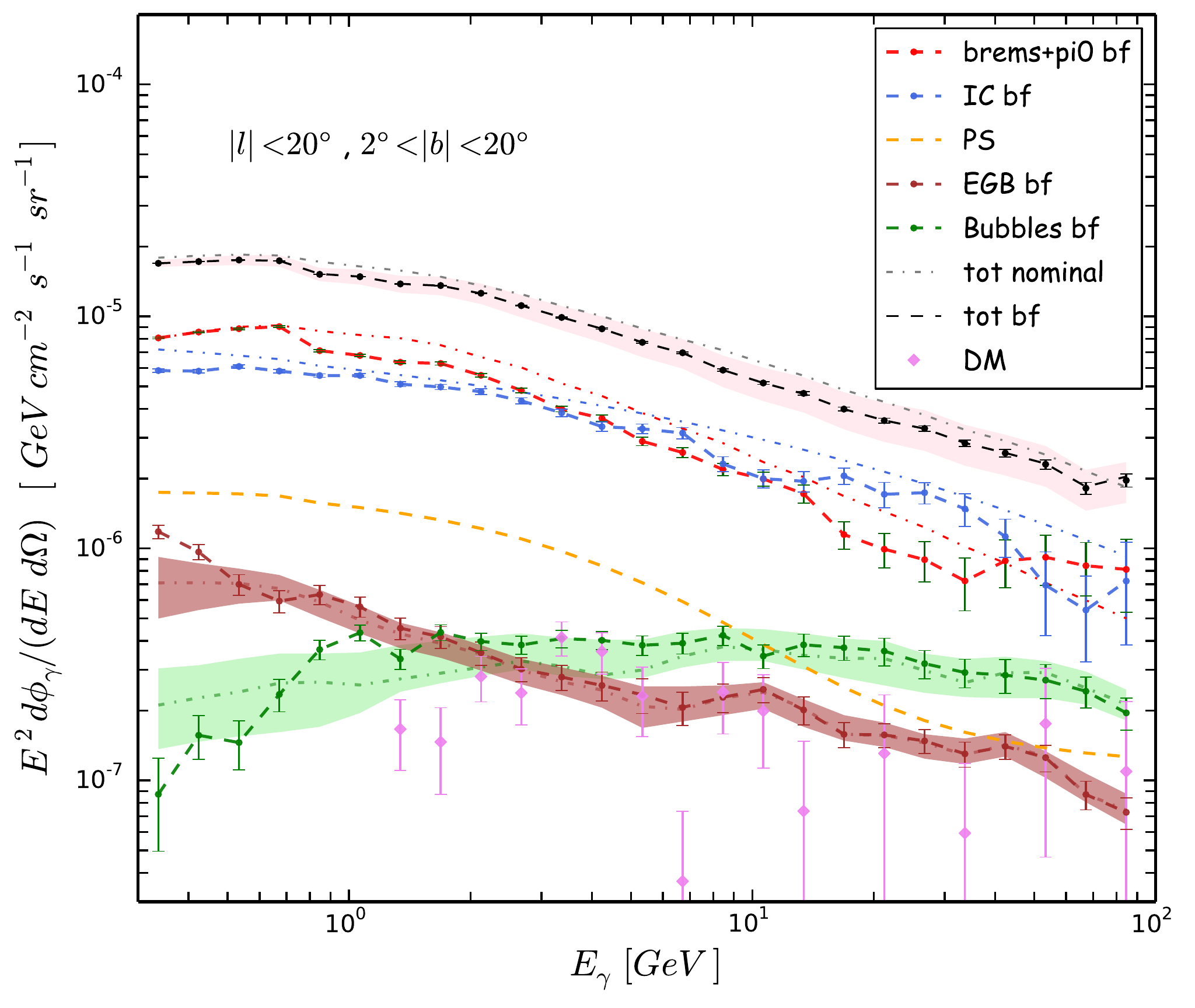}
\endminipage\\
\caption{ \textit{ Spectrum of the various contributions to the total $\gamma$-ray flux, pre- and post-template-fitting, compared to the Fermi-LAT data in the ROI characterized by $|l| < 20^{\circ}$, and $2^{\circ} < |b| < 20^{\circ}$. The violet band superimposed to the data represents the systematic uncertainty.
Left panel: Model A+DM. Right panel: Model A+spike+DM.
For Inverse Compton (light blue), $\pi^0$+Bremsstrahlung (red), isotropic background emission (dark red) and Fermi bubbles (green) dot-dashed lines show the nominal spectrum (pre-fitting) while points and dashed lines are the post-fitting values.
Uncertainties bands and central values for the isotropic background template (dark red) and Fermi bubbles (green) are taken respectively from~\cite{Ackermann:2014usa} and ~\cite{Fermi-LAT:2014sfa}.
Magenta diamonds: DM contribution. The point-source template (orange dashed line) is not touched by the fit. 
The error bars on the templates are obtained from the fitting procedure.
The black dashed (dot-dashed) line is the post (pre) template-fitting total spectrum.
}}
\label{fig:spectrum}
\end{figure}
\begin{figure}[!htb]
\minipage{0.33\textwidth}
  \includegraphics[width=1.\linewidth]{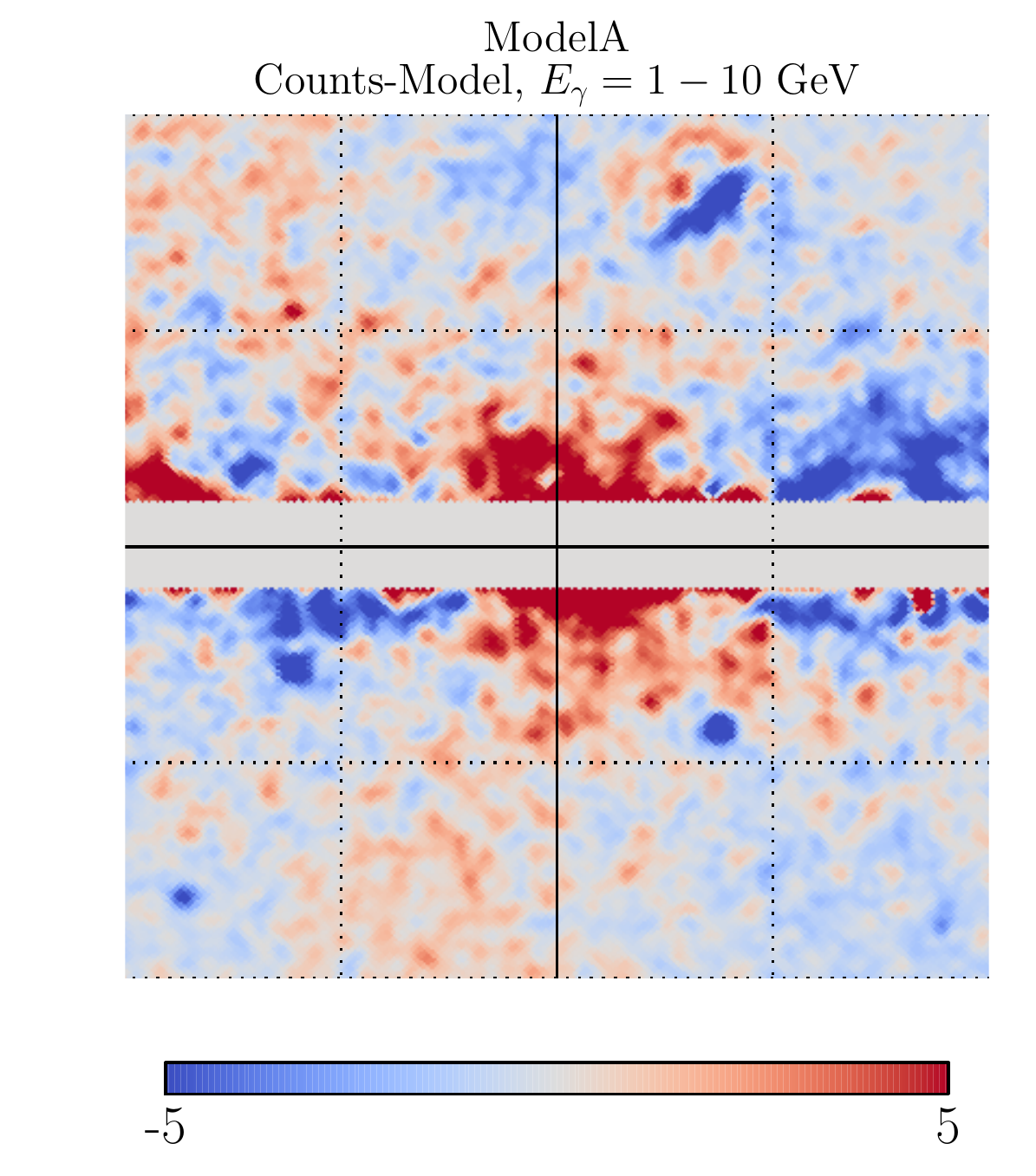}
\endminipage\hfill
\minipage{0.33\textwidth}
  \includegraphics[width=1.\linewidth]{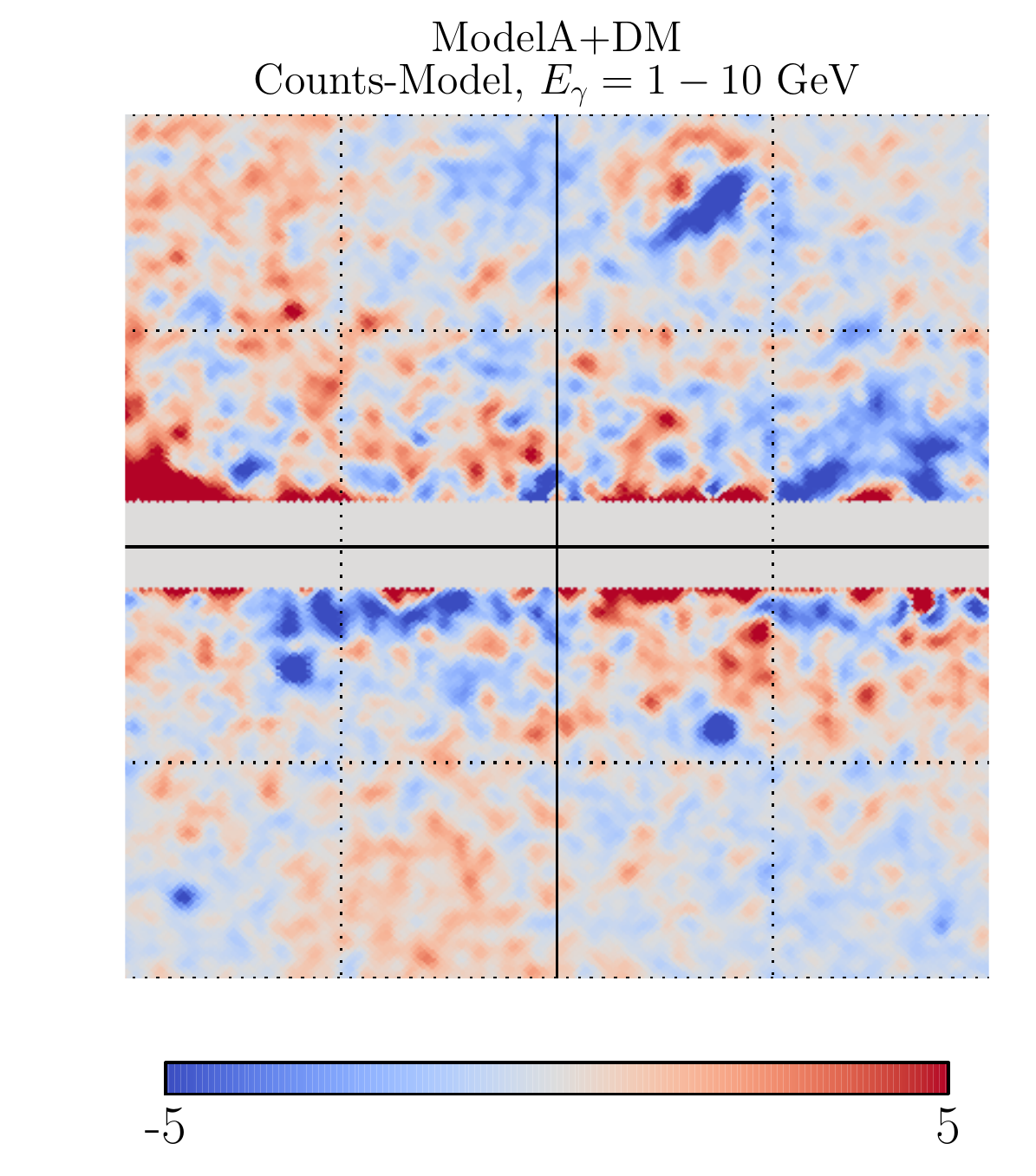}
\endminipage\hfill
\minipage{0.33\textwidth}
  \includegraphics[width=1.\linewidth]{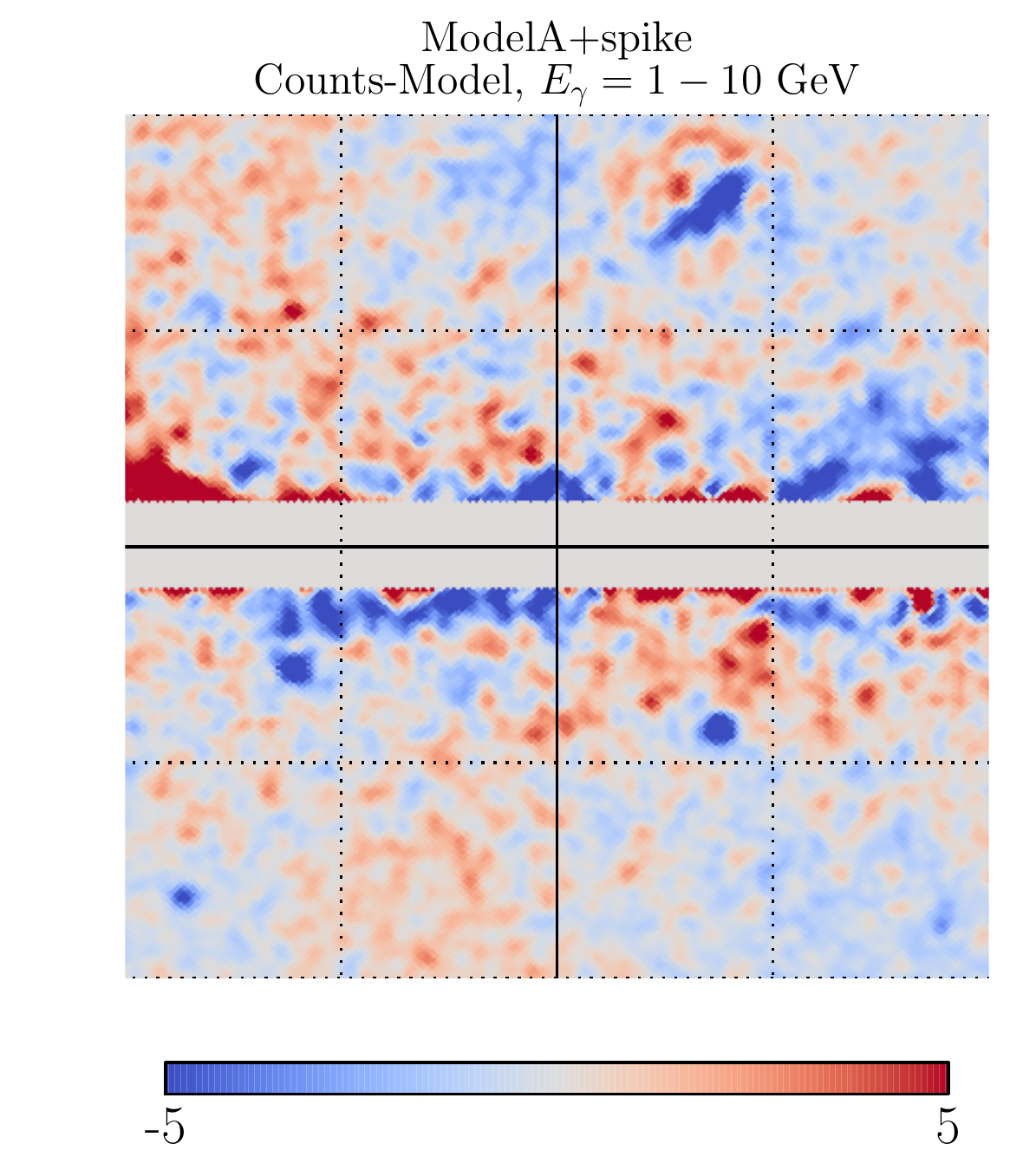}
\endminipage
 \caption{\small \textit{
Residual counts obtained for the Model A (left panel,  without the inclusion of DM template), 
for the Model A + DM (central panel), and for the reference model described in section~\ref{sec:ReferenceCase} (right panel, without the inclusion of DM template). See text in section~\ref{sec:EnergySpectrum} for a detailed discussion.
  }}
 \label{fig:residuals}
\end{figure}

\begin{figure}[!htb!]
\minipage{0.5\textwidth}
  \includegraphics[width=1.\linewidth]{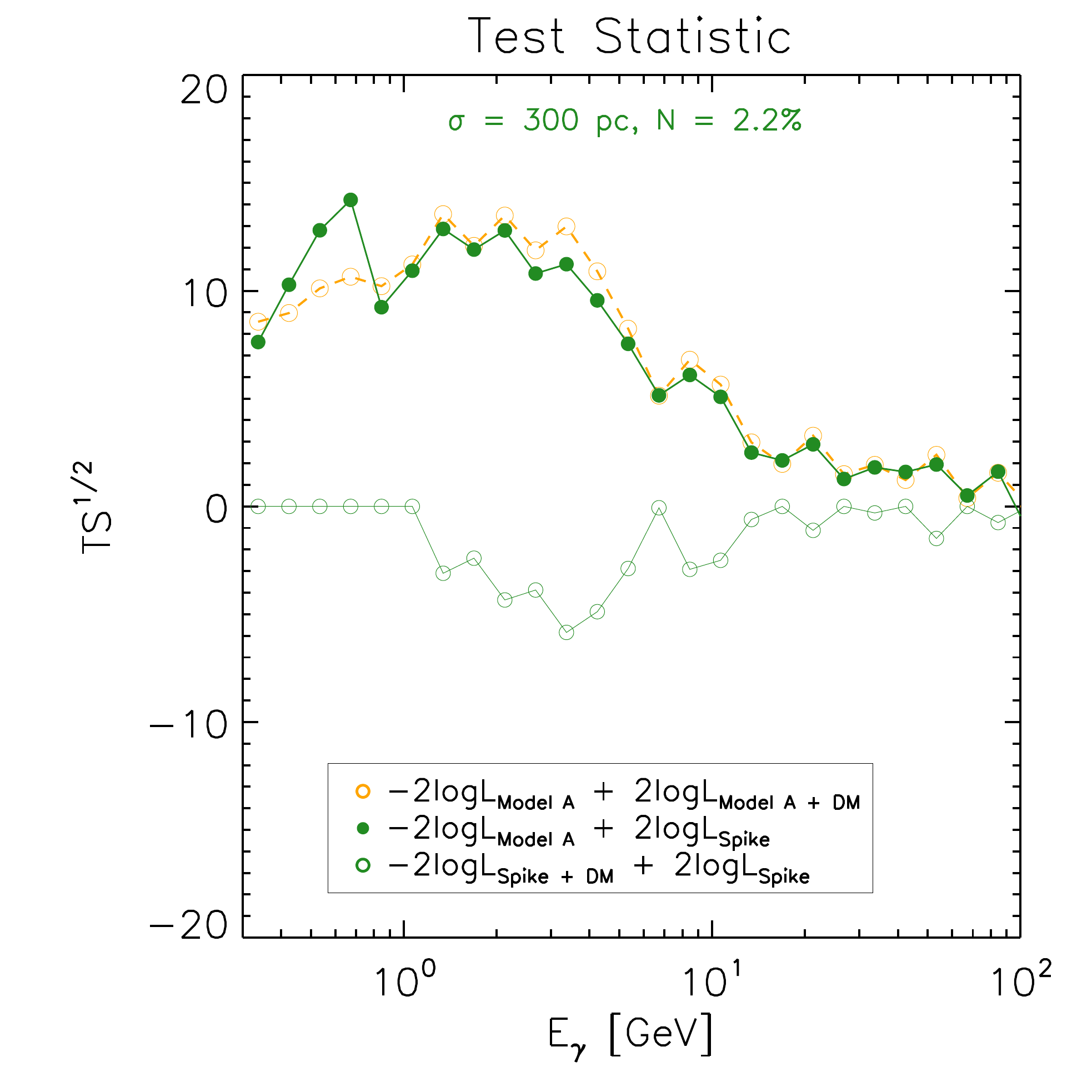}
\endminipage\hfill
\minipage{0.5\textwidth}
  \includegraphics[width=1.\linewidth]{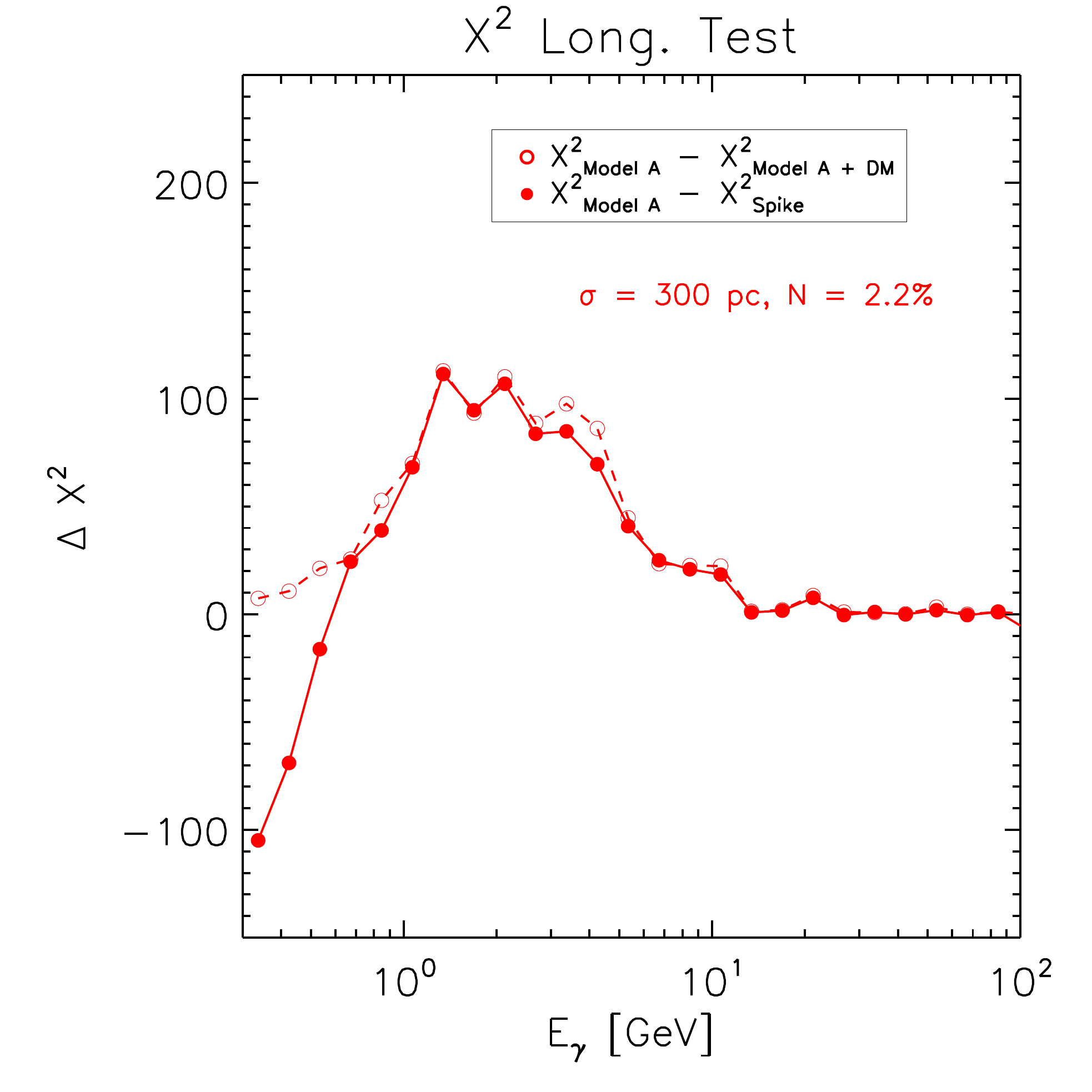}
\endminipage\\
\minipage{0.5\textwidth}
  \includegraphics[width=1.\linewidth]{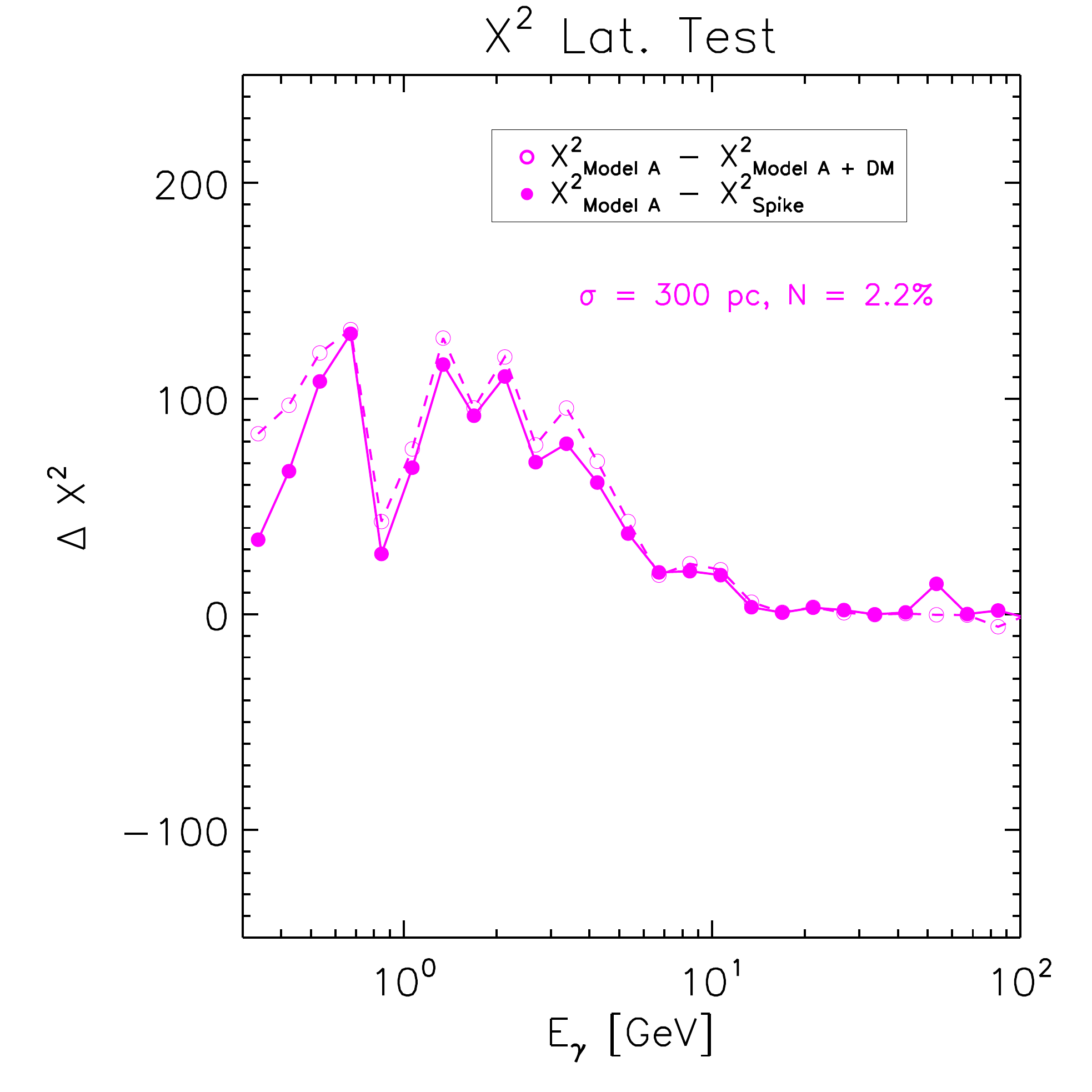}
\endminipage\hfill
\minipage{0.5\textwidth}
  \includegraphics[width=1.\linewidth]{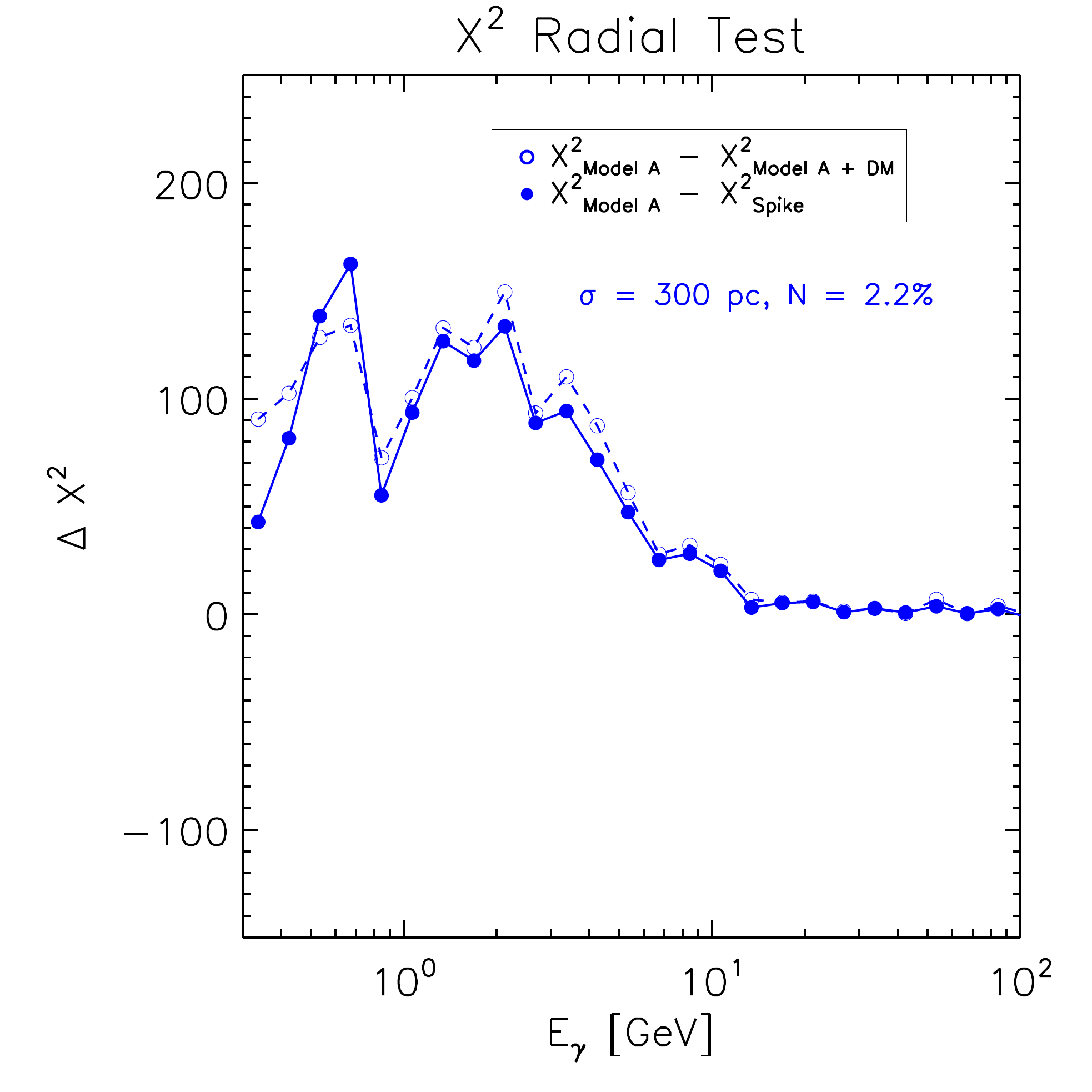}
\endminipage\\
\caption{\small \textit{ {\bf Top left panel}. We compare the test-statistic (${\rm TS} = -2\Delta\log\mathcal{L},$  we show the square-root of TS) of the models we consider; a positive difference between two models means that the second model performs better. Yellow filled circles: $-2\log\mathcal{L}_{\rm Model\,A}+2\log\mathcal{L}_{\rm Model\,A + DM}$. Green filled circles: $-2\log\mathcal{L}_{\rm Model\,A}+2\log\mathcal{L}_{\rm Spike}$. Green empty circles: $-2\log\mathcal{L}_{\rm Spike + DM}+2\log\mathcal{L}_{\rm Spike}$.  {\bf Top right panel}. We compare the $\chi^2$ of the longitude profiles for the same models. Filled circles: $\chi^2_{\rm Model\,A}- \chi^2_{\rm Spike}$; empty circles: $\chi^2_{\rm Model\,A}- \chi^2_{\rm Model\,A + DM}$. {\bf Bottom panels}. The same as the top right panel, for latitude and radial profiles. }}
\label{fig:likelihood}
\end{figure}

\begin{figure}[!htb]
\minipage{0.33\textwidth}
  \includegraphics[width=1.\linewidth]{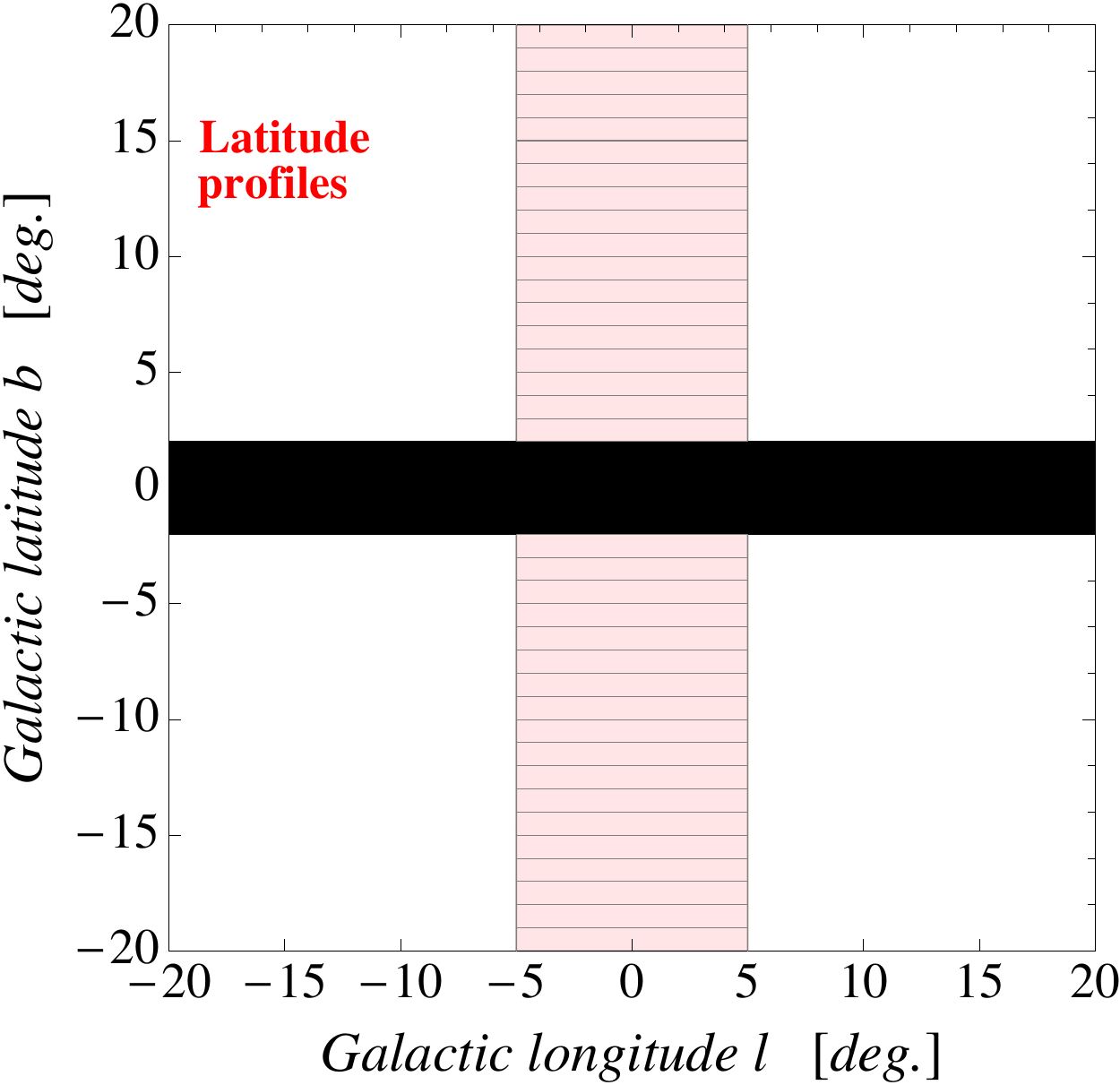}
\endminipage\hfill
\minipage{0.33\textwidth}
  \includegraphics[width=1.\linewidth]{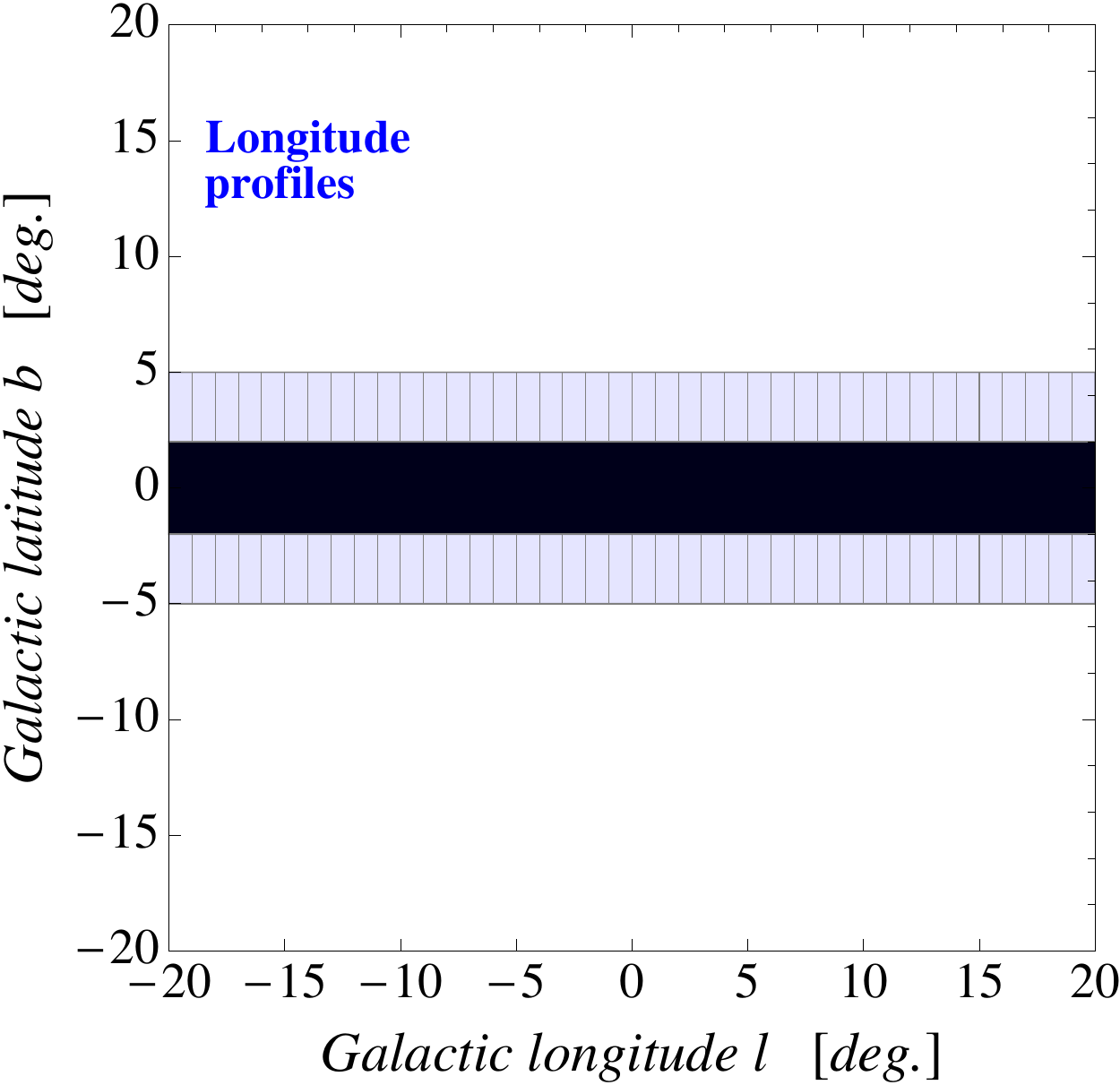}
\endminipage\hfill
\minipage{0.33\textwidth}
  \includegraphics[width=1.\linewidth]{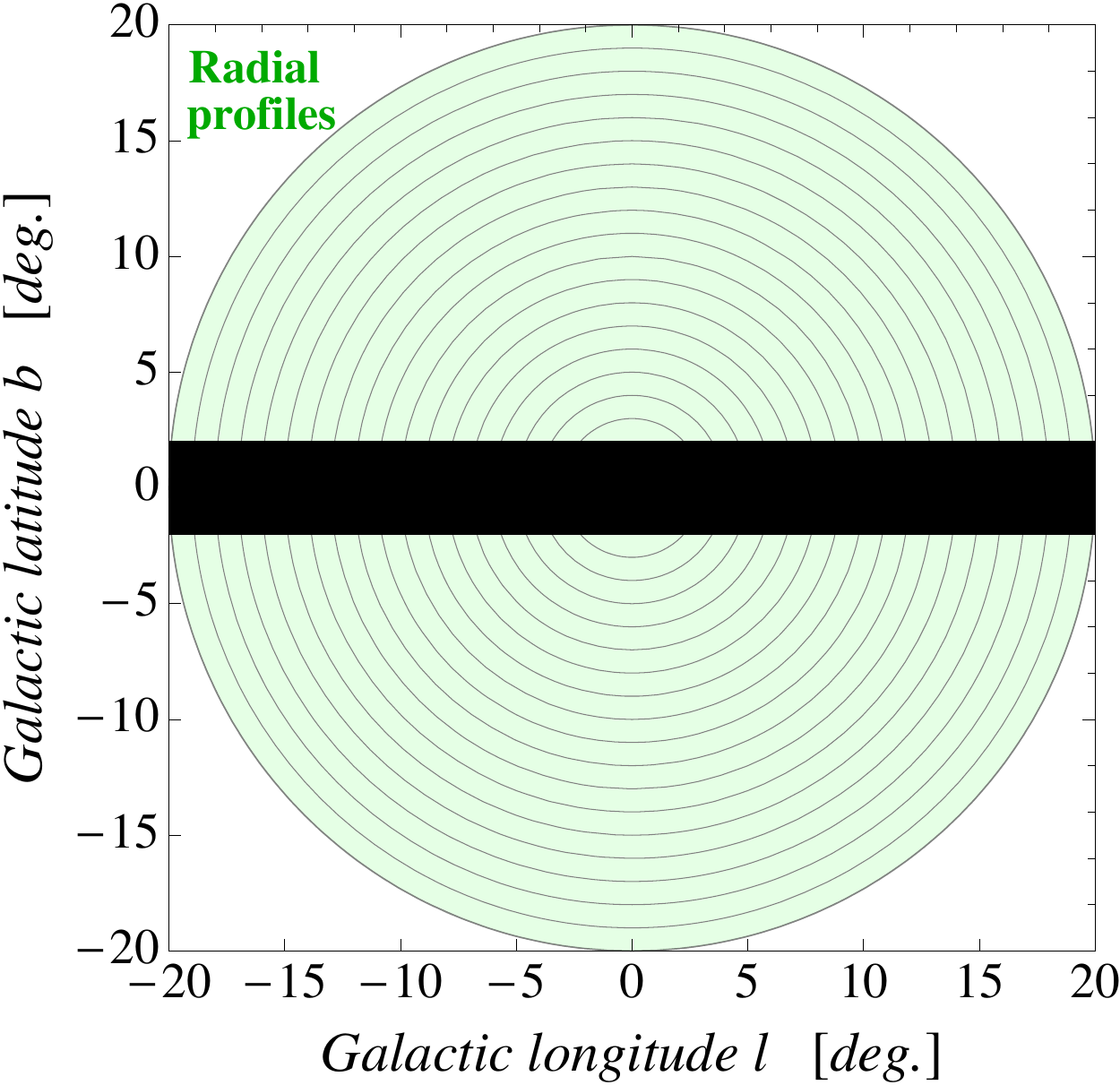}
\endminipage
 \caption{\small \textit{
 Control regions used for the computation of the $\gamma$-ray profiles in section~\ref{sec:Profiles}. See text for details.
  }}
 \label{fig:lProfiles}
\end{figure}

\begin{figure}[!htb]
 \begin{center}
\fbox{\footnotesize \textbf{{\color{red}{$\chi^2$ comparison: latitude profiles}}}}
\end{center}
\vspace{0.cm}
\minipage{0.33\textwidth}
  \includegraphics[width=1.1\linewidth]{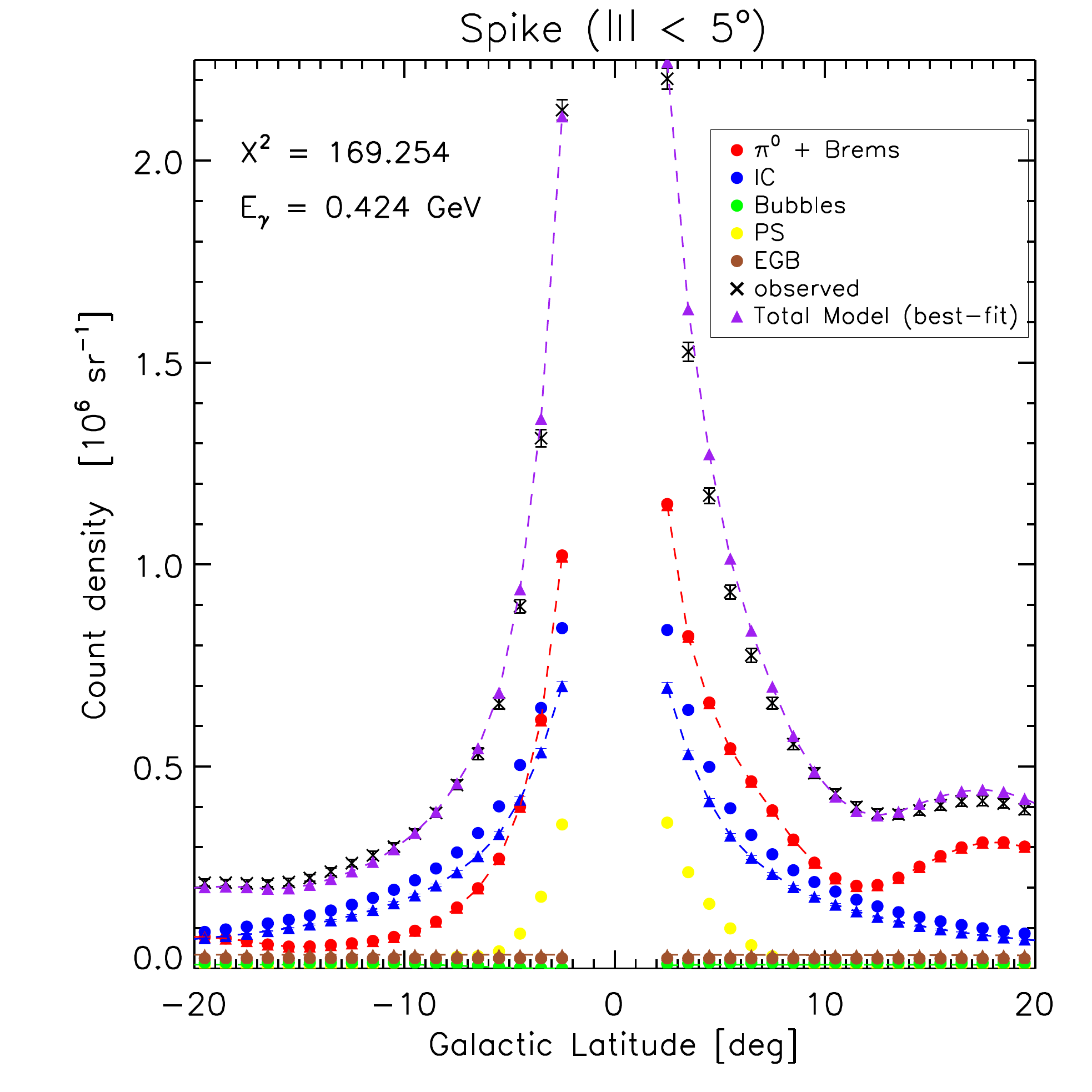}
\endminipage\hfill
\minipage{0.33\textwidth}
  \includegraphics[width=1.1\linewidth]{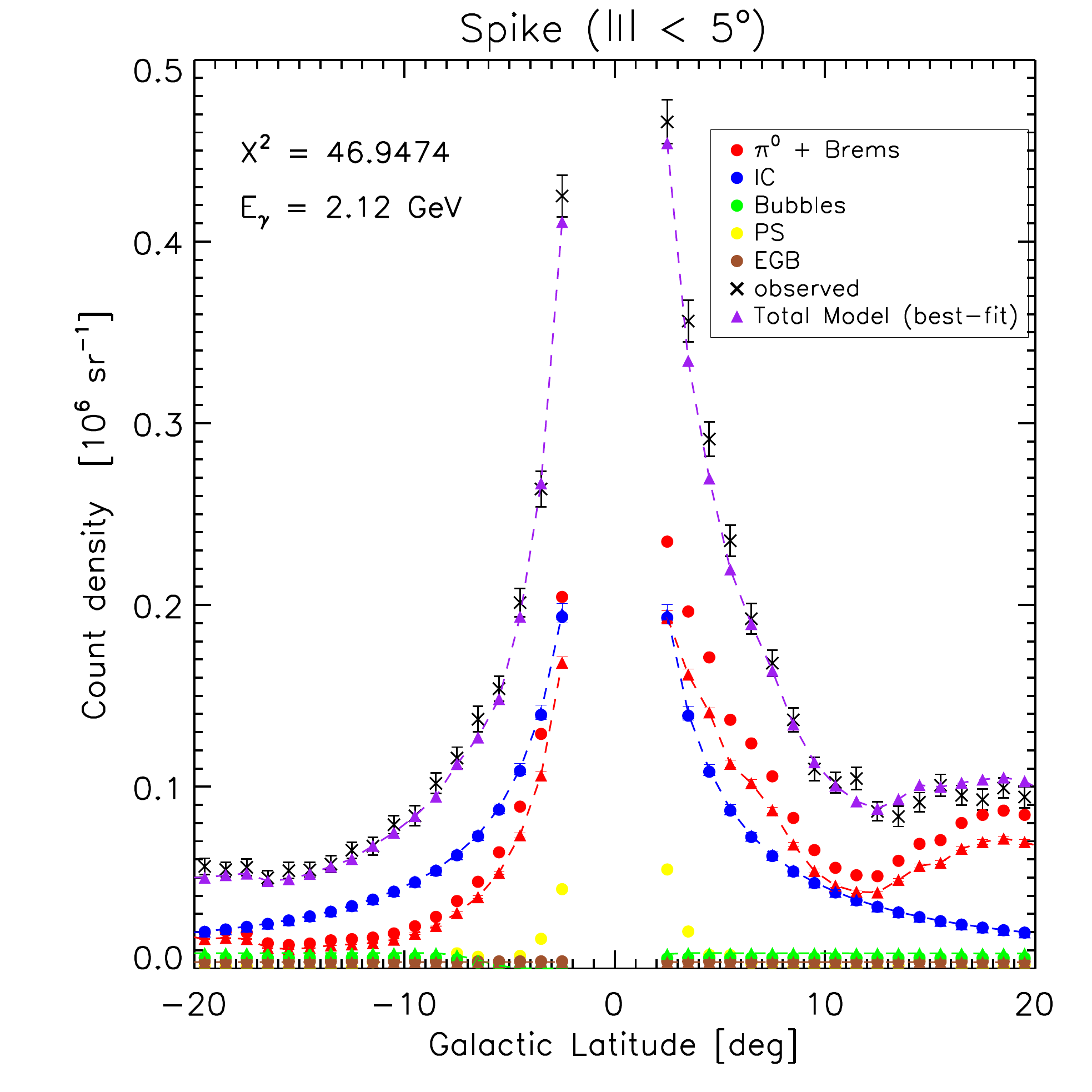}
\endminipage\hfill
\minipage{0.33\textwidth}
  \includegraphics[width=1.1\linewidth]{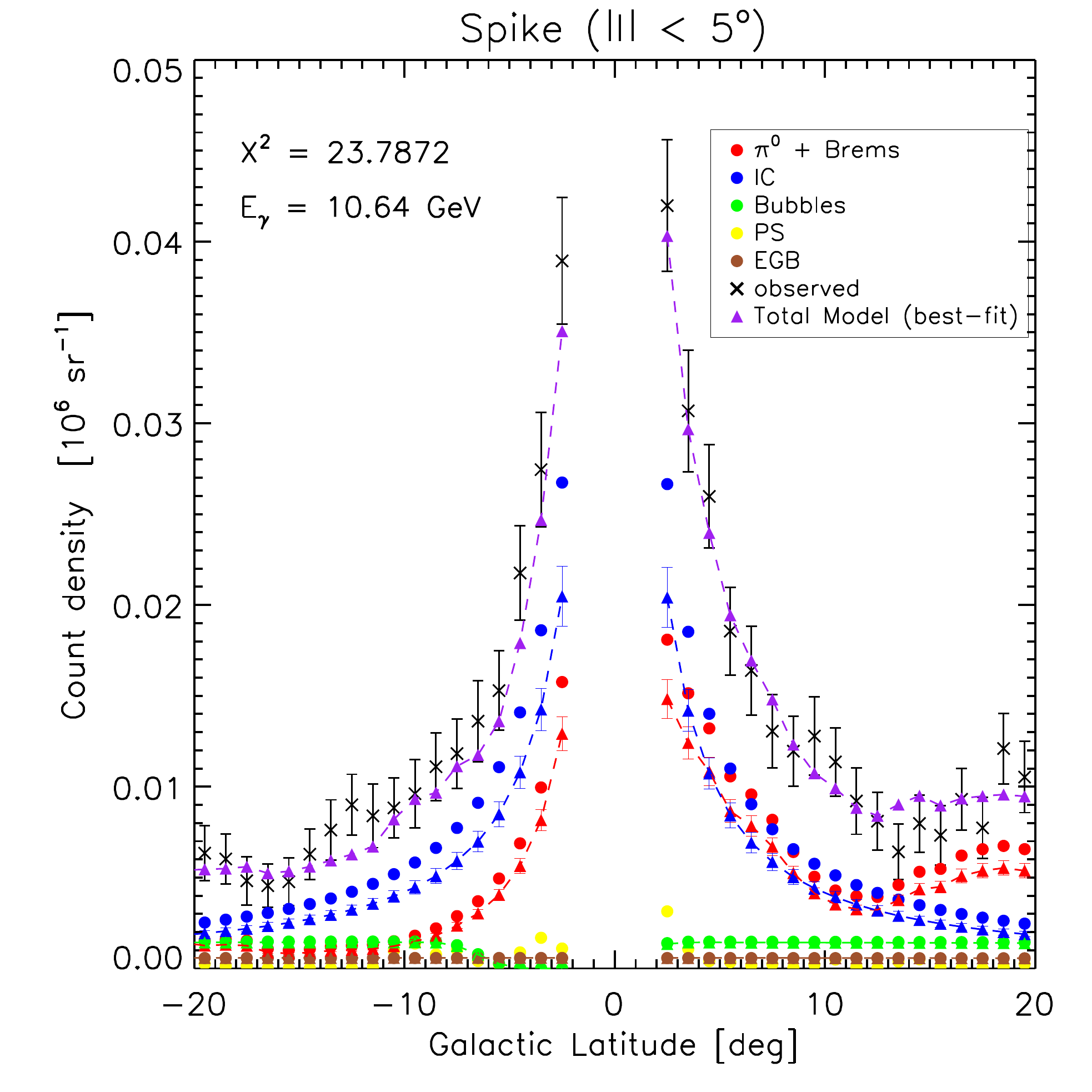}
\endminipage \vspace{.3 cm}\\
\minipage{0.33\textwidth}
  \includegraphics[width=1.1\linewidth]{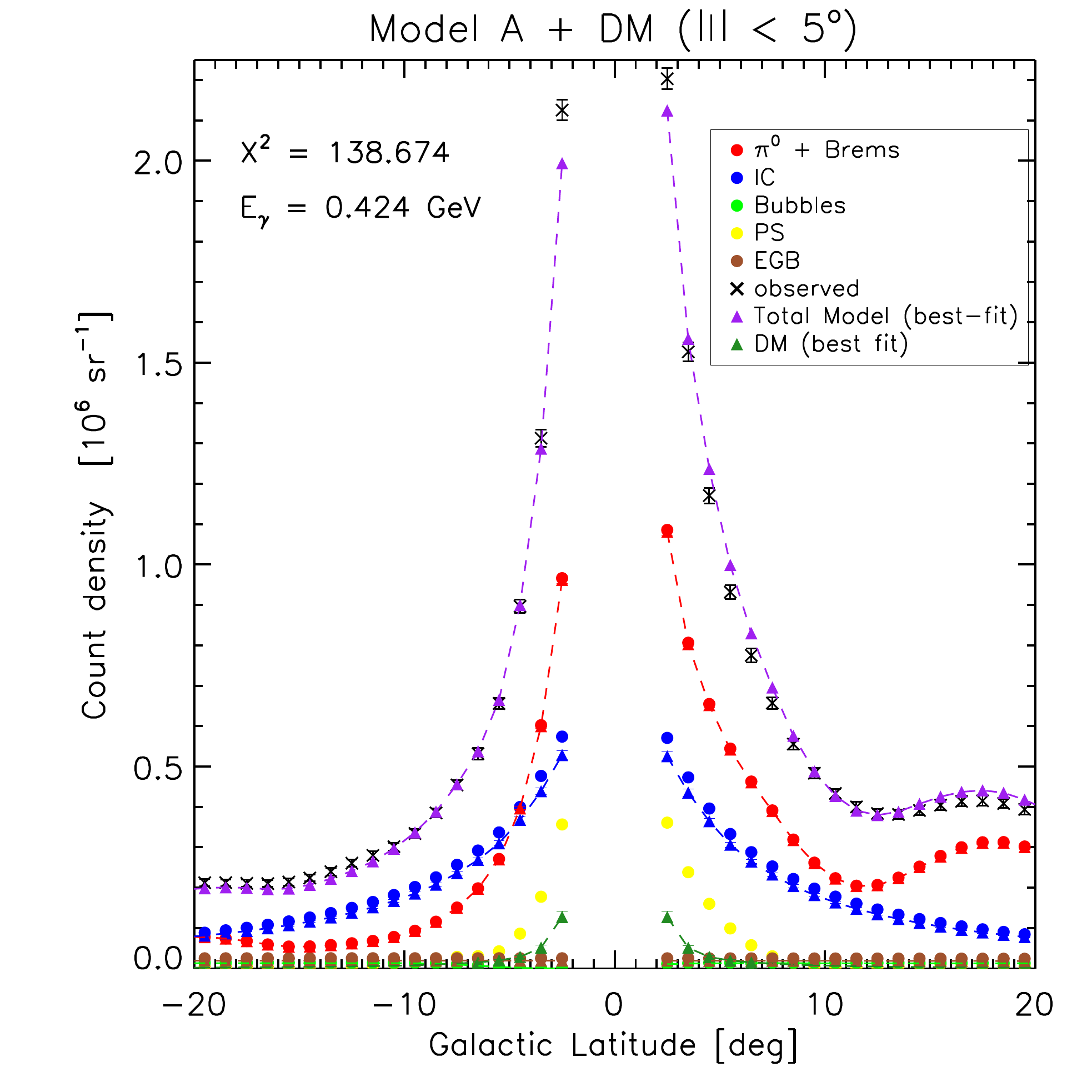}
\endminipage\hfill
\minipage{0.33\textwidth}
  \includegraphics[width=1.1\linewidth]{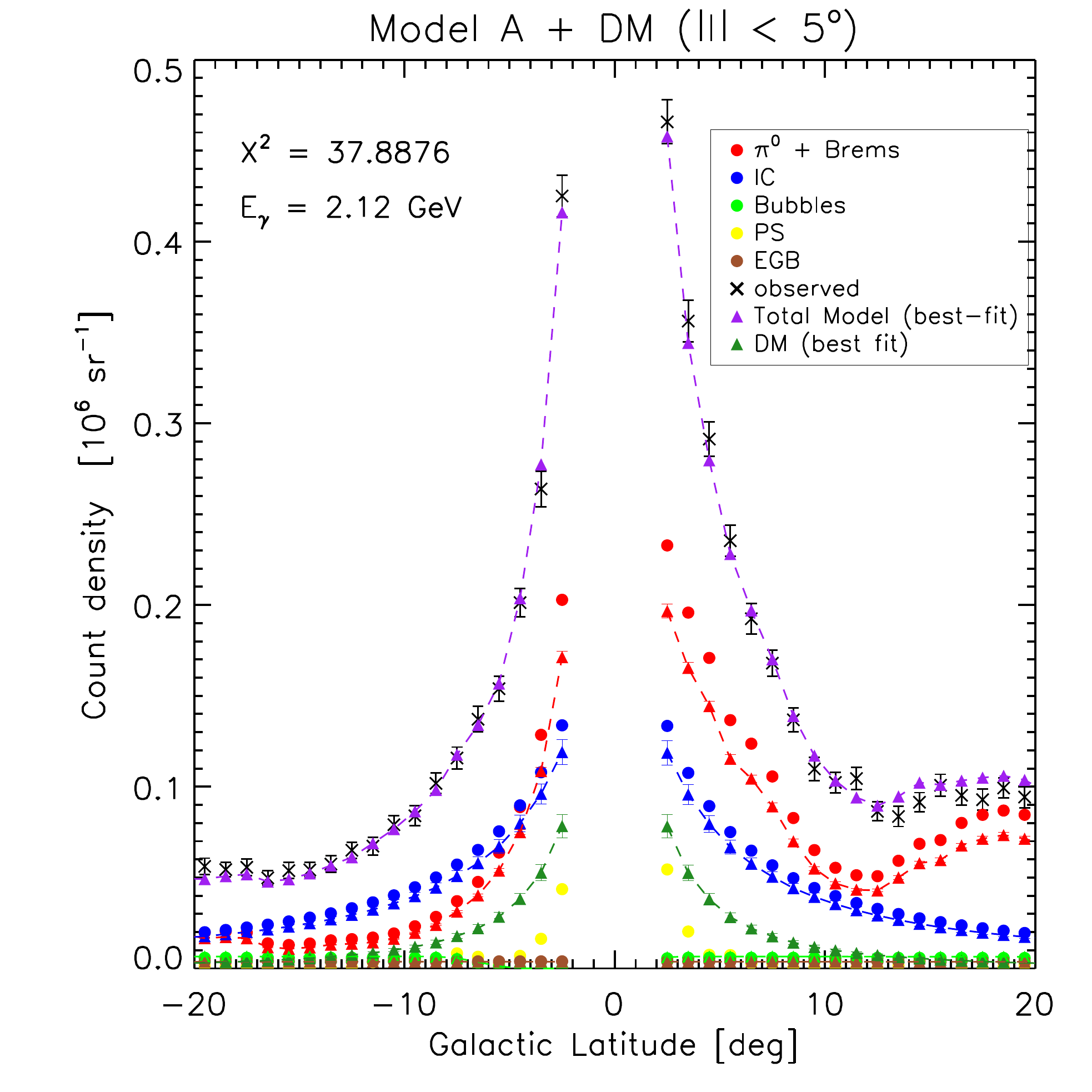}
\endminipage\hfill
\minipage{0.33\textwidth}
  \includegraphics[width=1.1\linewidth]{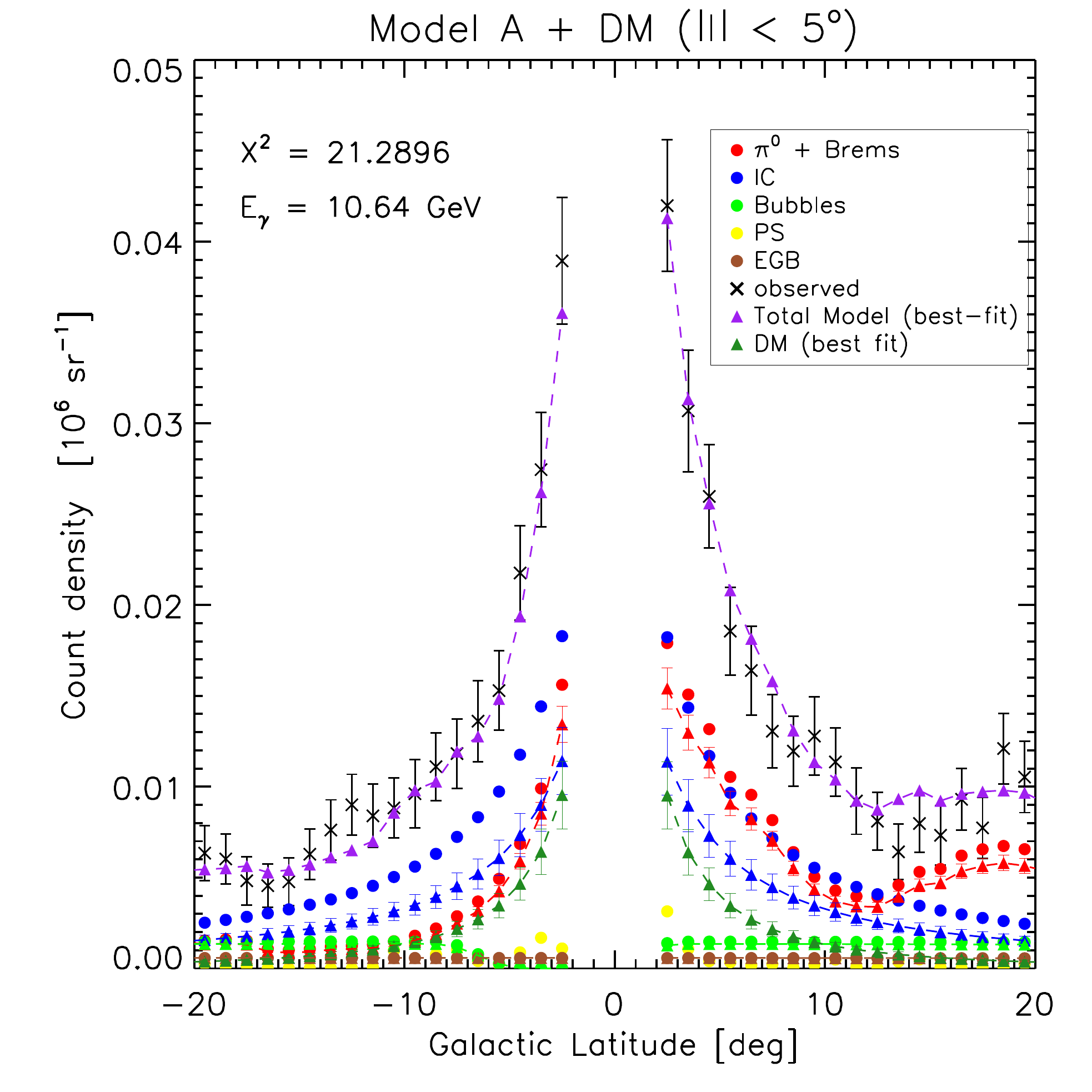}
\endminipage\\
 \caption{\small \textit{ {\bf Top row.} The latitudinal profiles for our best case at several energies. From left to right, $400$ MeV, $2$ GeV, $10$ GeV. Red circles (triangles): $\pi^0$+Bremsstrahlung contribution, pre- and post-fitting. Blue circles (triangles): IC contribution, pre- and post-fitting. Magenta triangles: total model, to be compared with Fermi-LATdata. {\bf Bottom row.} The same plots for Model A+DM (DM contribution in forest green triangles).}}
 \label{fig:lat_profiles}
\end{figure}

\begin{figure}[!htb]
 \begin{center}
\fbox{\footnotesize \textbf{{\color{blue}{$\chi^2$ comparison: longitude profiles}}}}
\end{center}
\vspace{0.cm}
\minipage{0.33\textwidth}
  \includegraphics[width=1.1\linewidth]{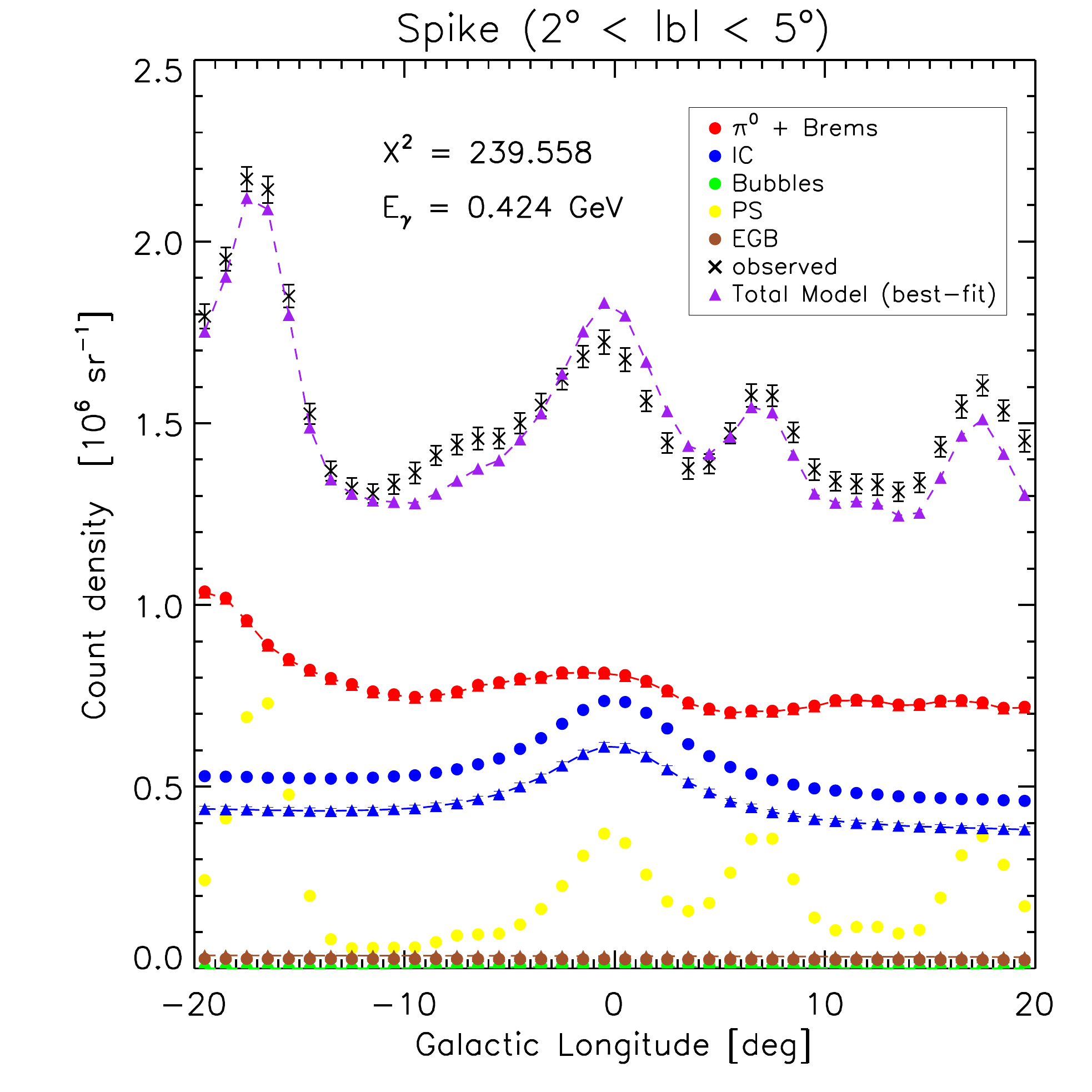}
\endminipage\hfill
\minipage{0.33\textwidth}
  \includegraphics[width=1.1\linewidth]{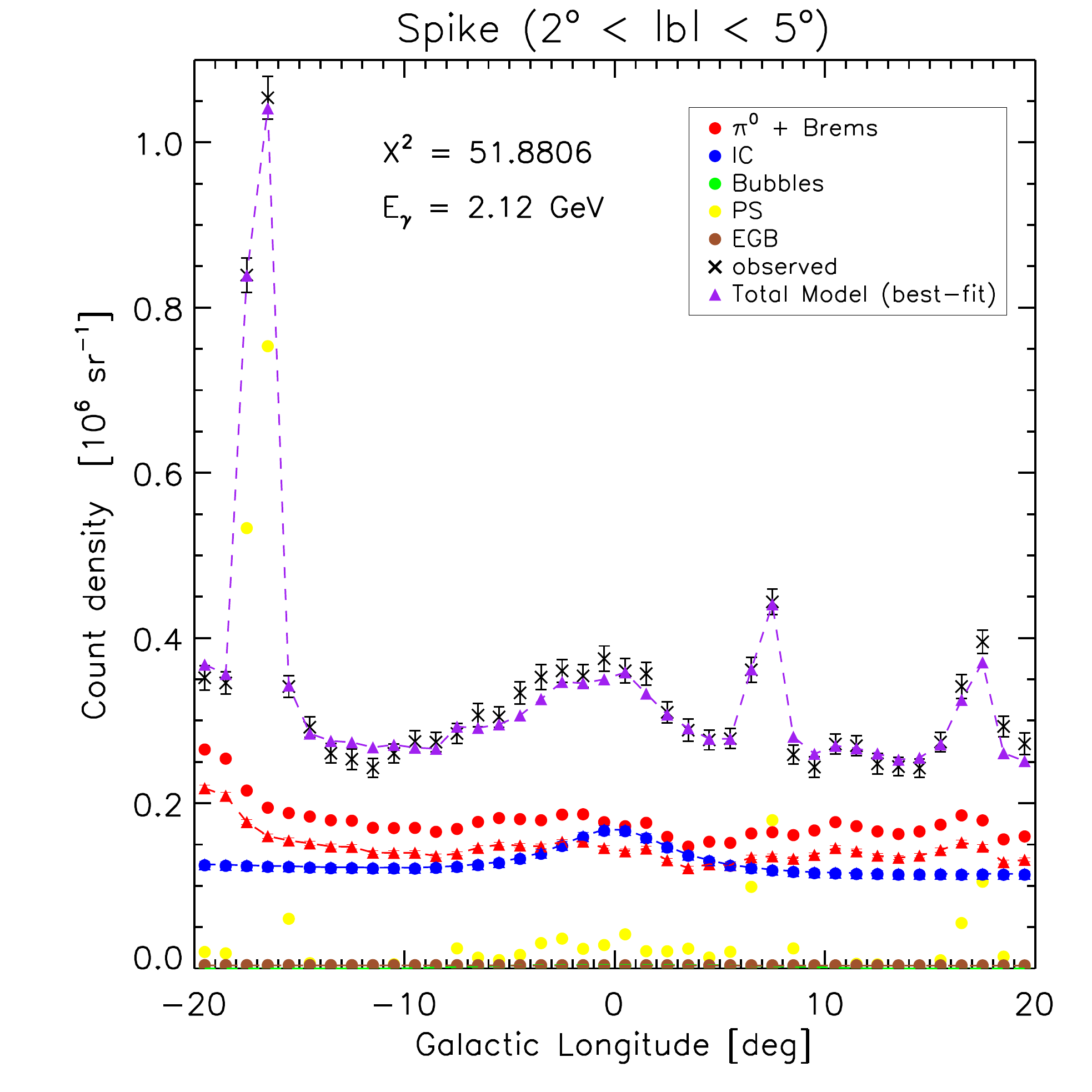}
\endminipage\hfill
\minipage{0.33\textwidth}
  \includegraphics[width=1.1\linewidth]{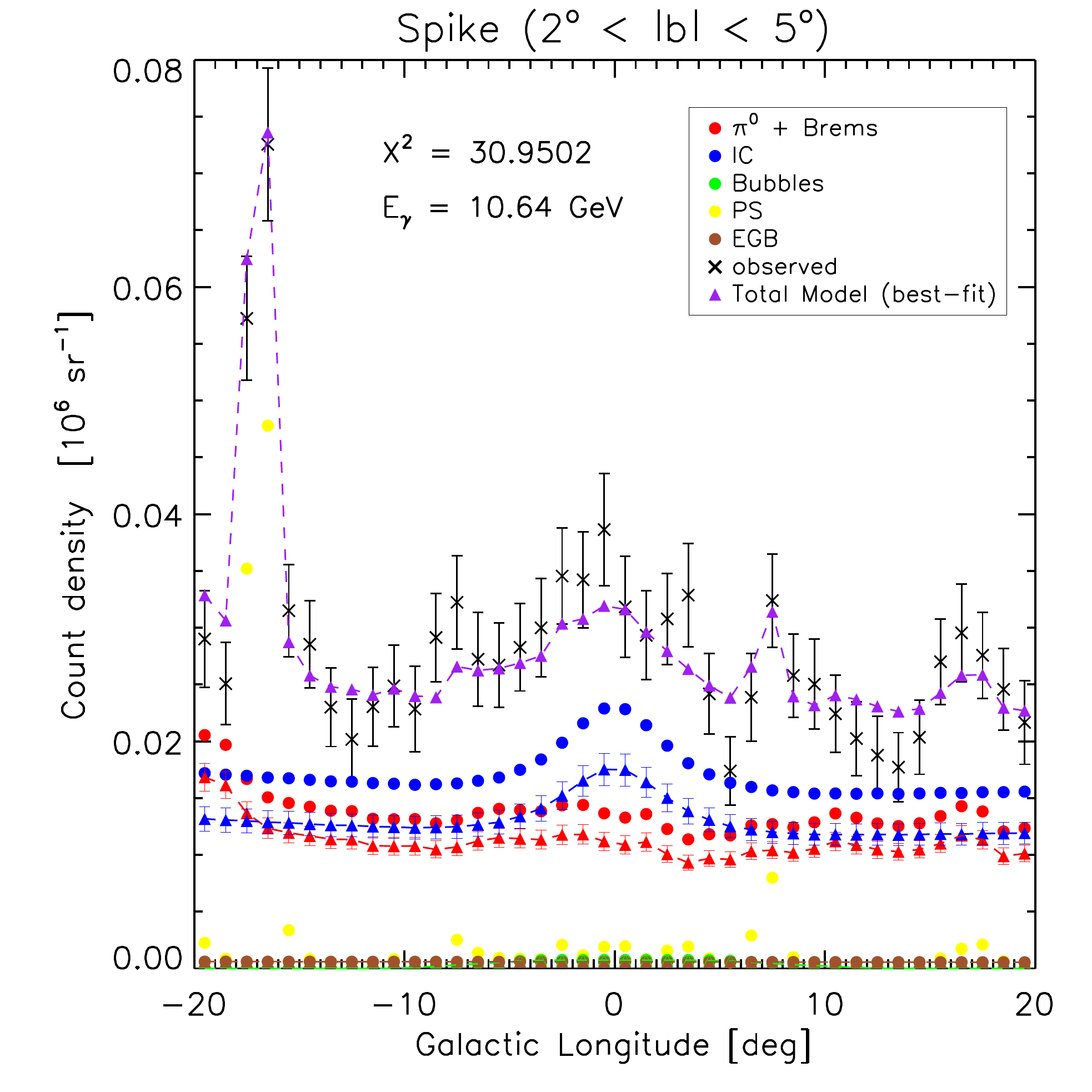}
\endminipage \vspace{.3 cm}\\
\minipage{0.33\textwidth}
  \includegraphics[width=1.1\linewidth]{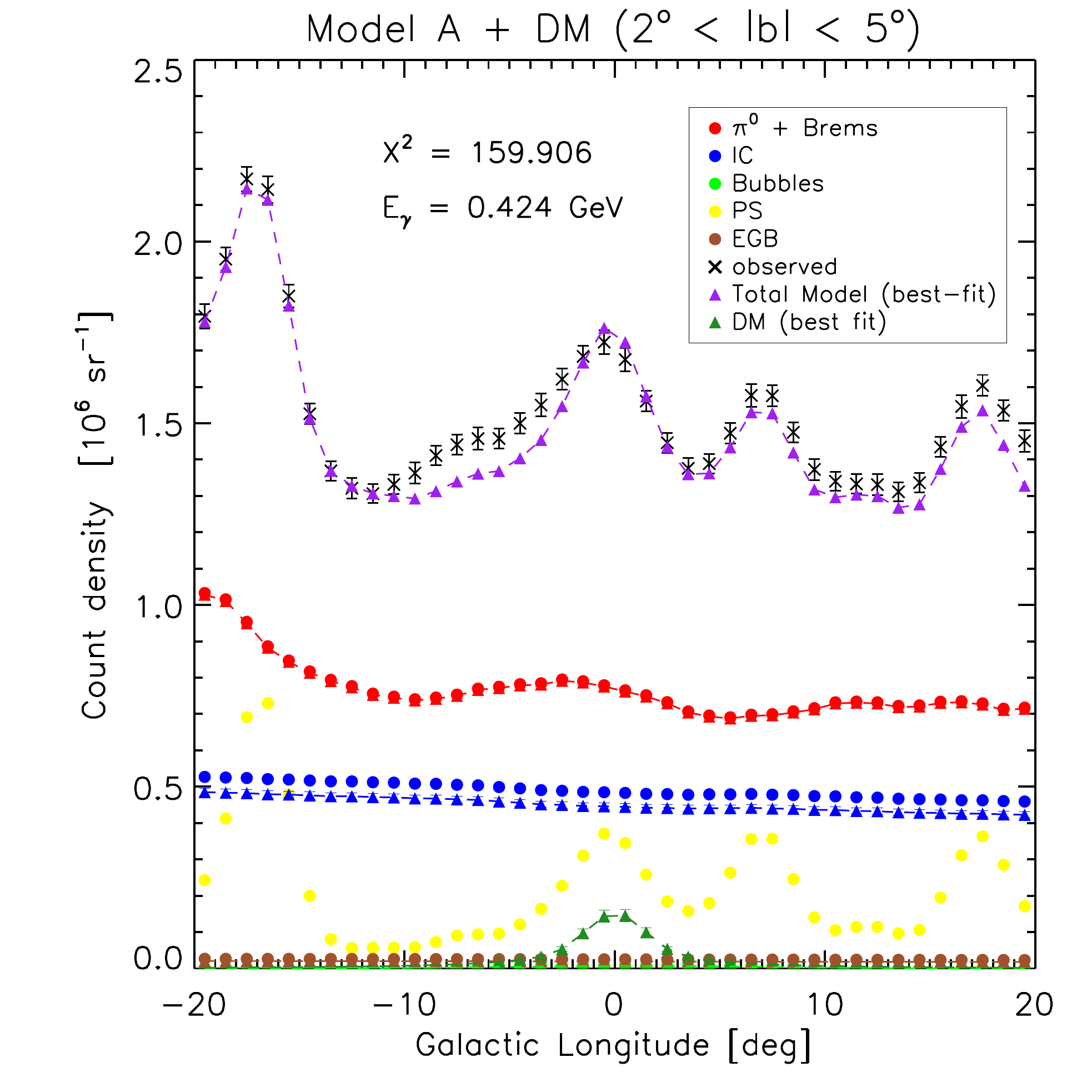}
\endminipage\hfill
\minipage{0.33\textwidth}
  \includegraphics[width=1.1\linewidth]{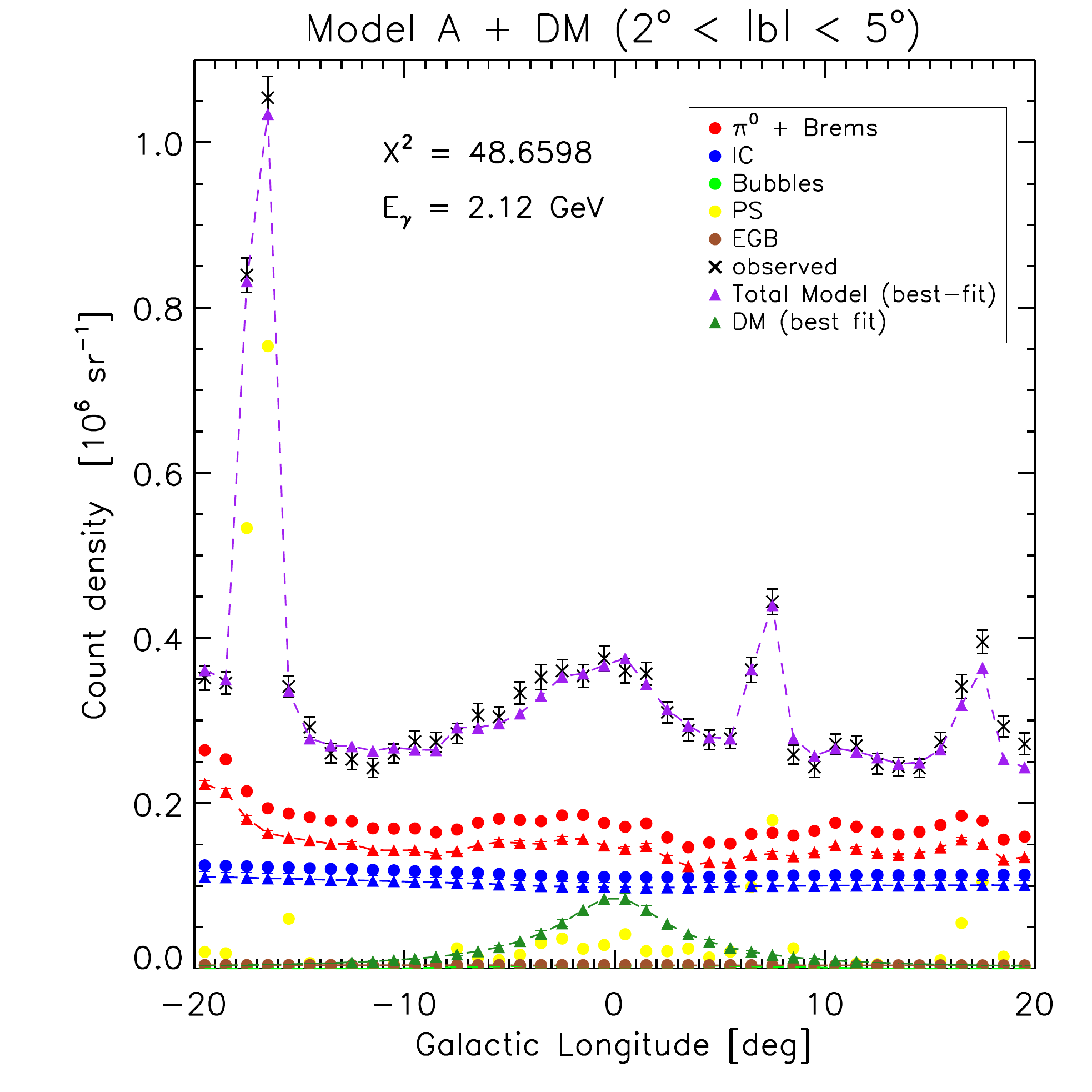}
\endminipage\hfill
\minipage{0.33\textwidth}
  \includegraphics[width=1.1\linewidth]{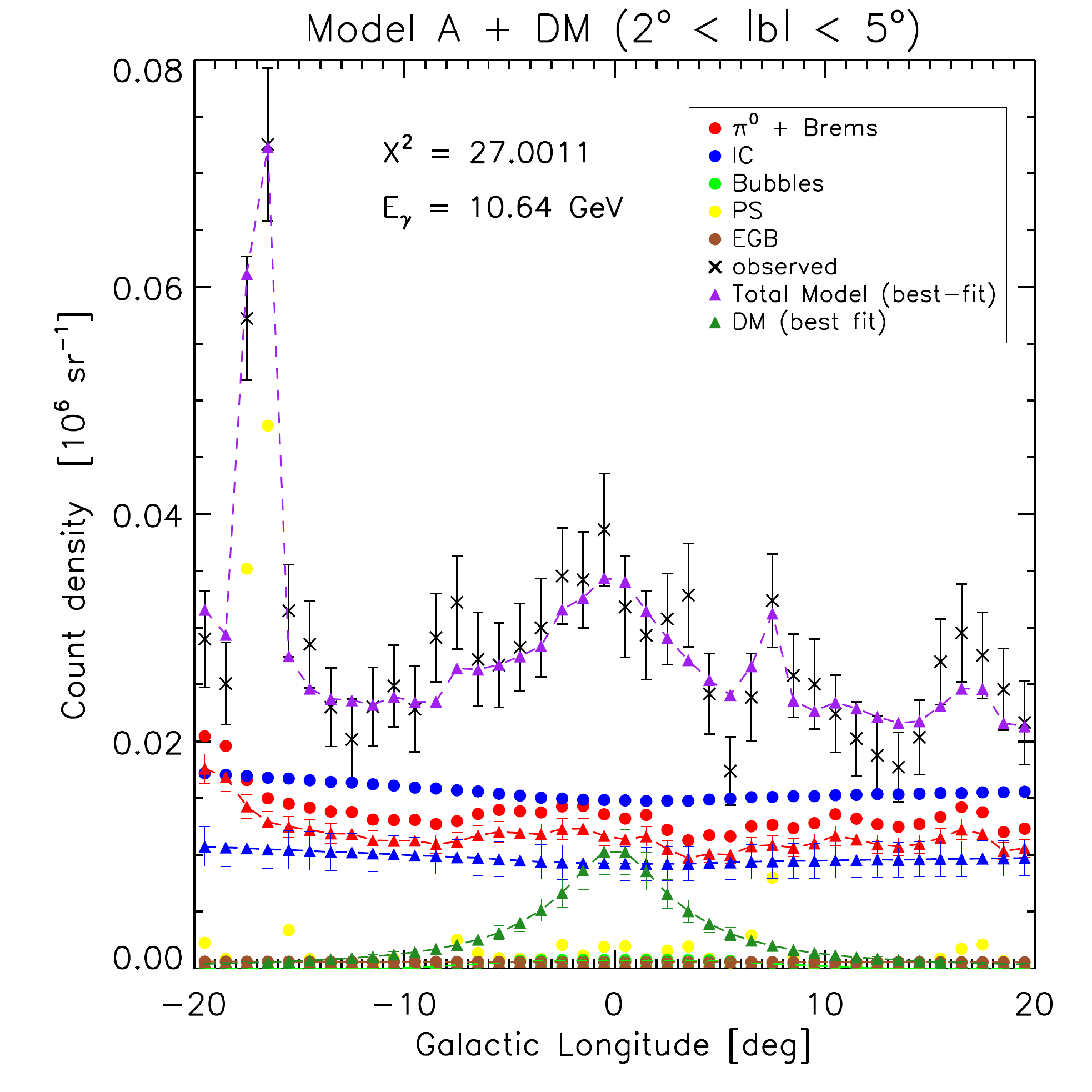}
\endminipage\\
 \caption{\small \textit{The same as fig. \ref{fig:lat_profiles} for the longitudinal profiles.}}
 \label{fig:long_profiles}
\end{figure}

\begin{figure}[!htb]
 \begin{center}
\fbox{\footnotesize \textbf{{\color{forestgreen}{$\chi^2$ comparison: radial profiles}}}}
\end{center}
\vspace{0.cm}
\minipage{0.33\textwidth}
  \includegraphics[width=1.1\linewidth]{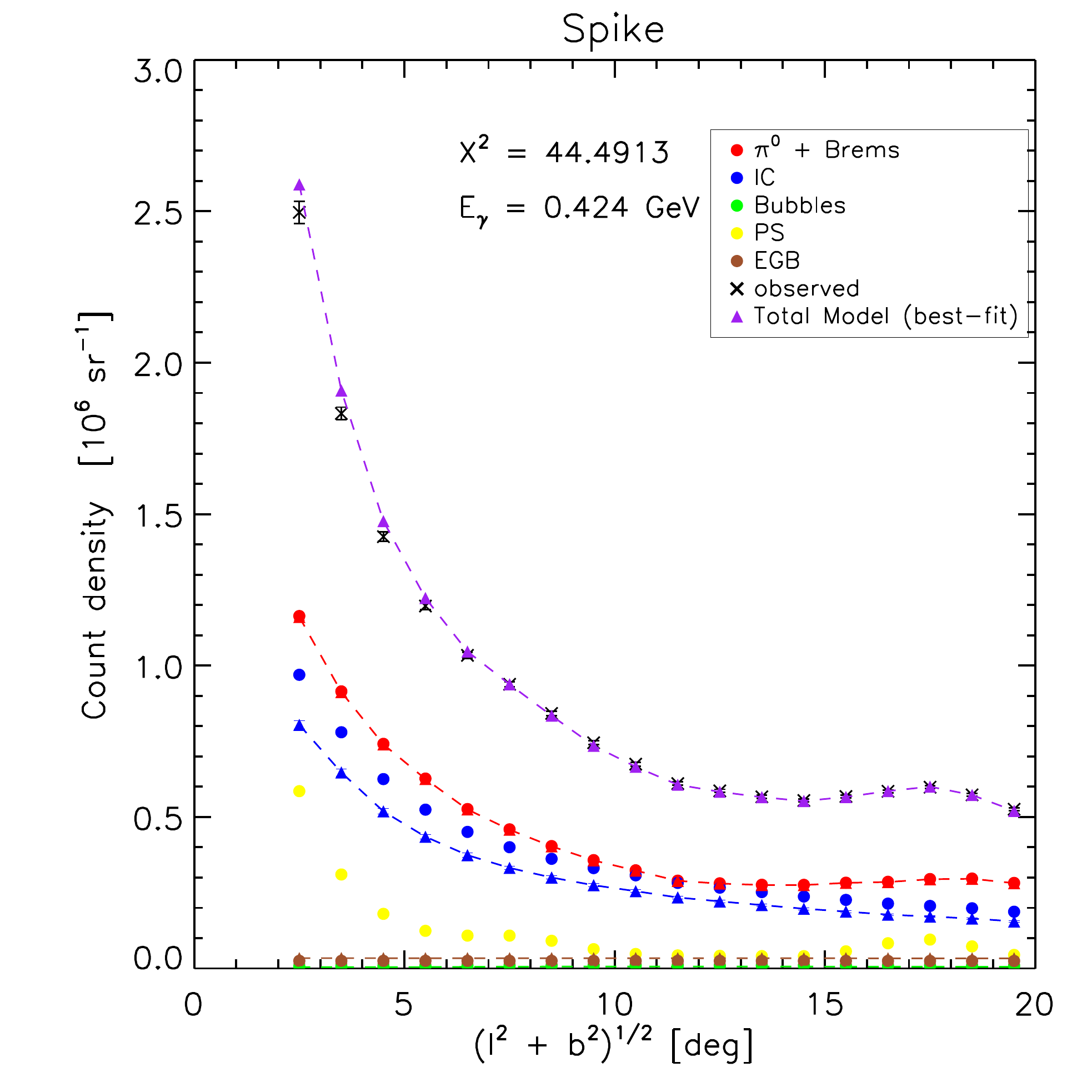}
\endminipage\hfill
\minipage{0.33\textwidth}
  \includegraphics[width=1.1\linewidth]{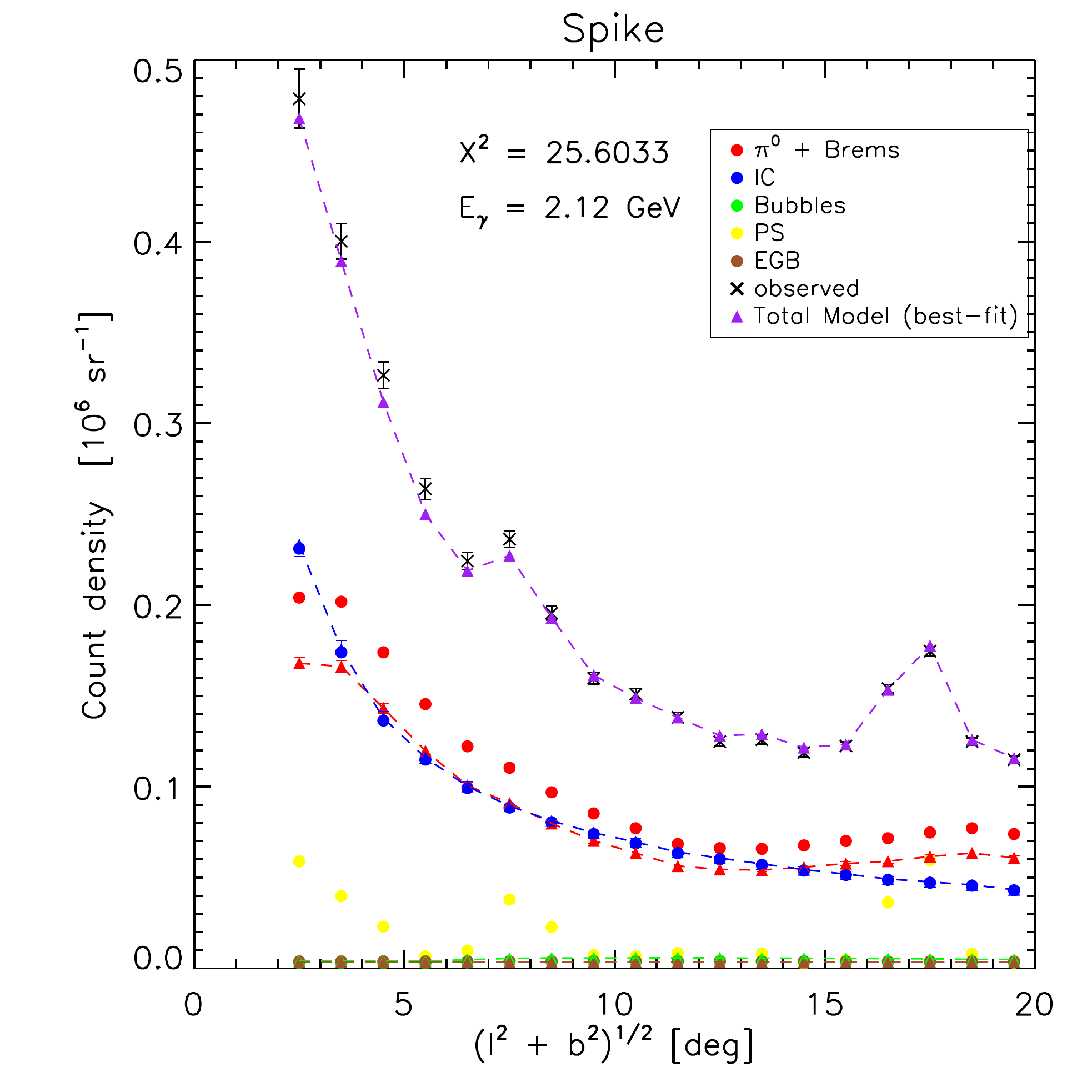}
\endminipage\hfill
\minipage{0.33\textwidth}
  \includegraphics[width=1.1\linewidth]{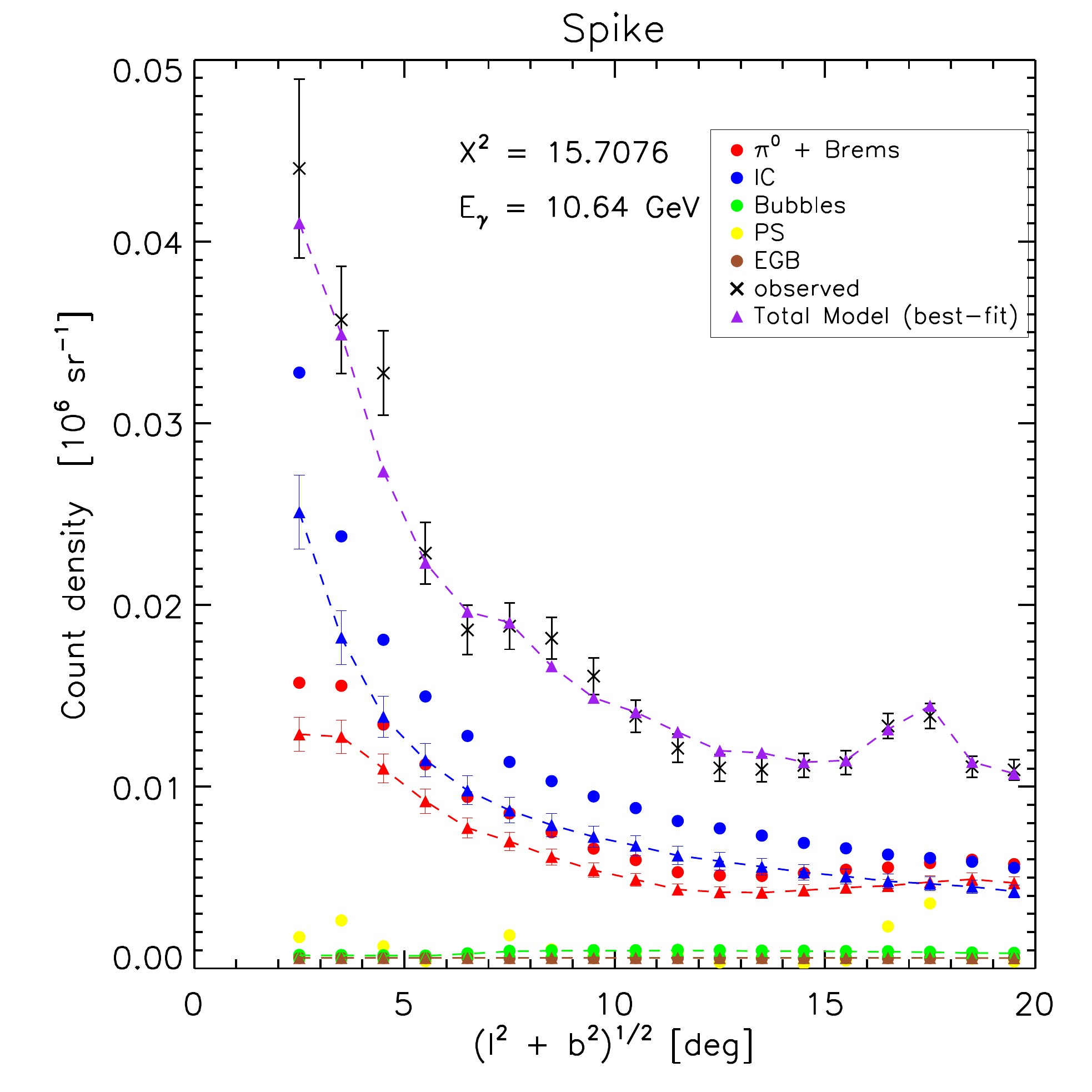}
\endminipage \vspace{.3 cm}\\
\minipage{0.33\textwidth}
  \includegraphics[width=1.1\linewidth]{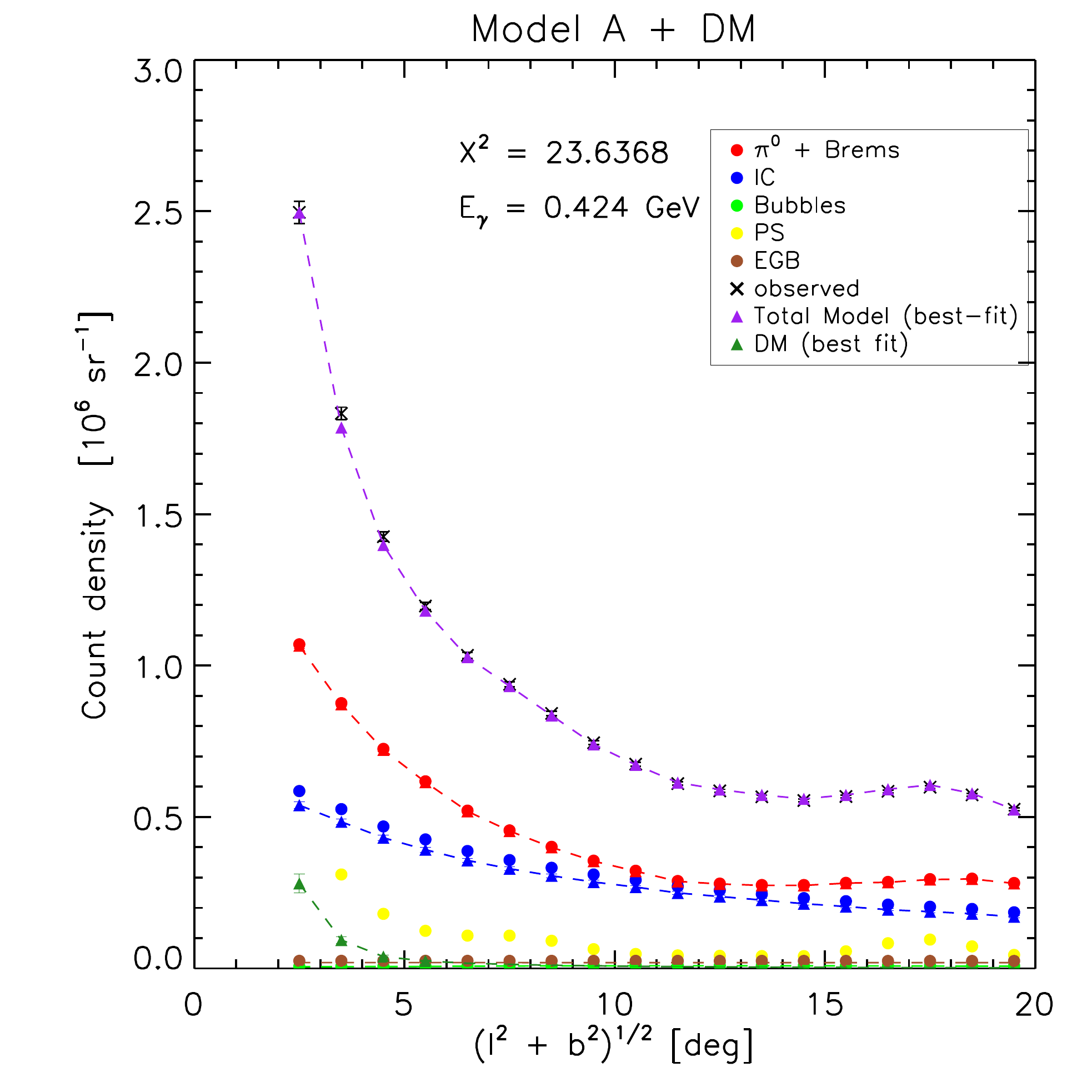}
\endminipage\hfill
\minipage{0.33\textwidth}
  \includegraphics[width=1.1\linewidth]{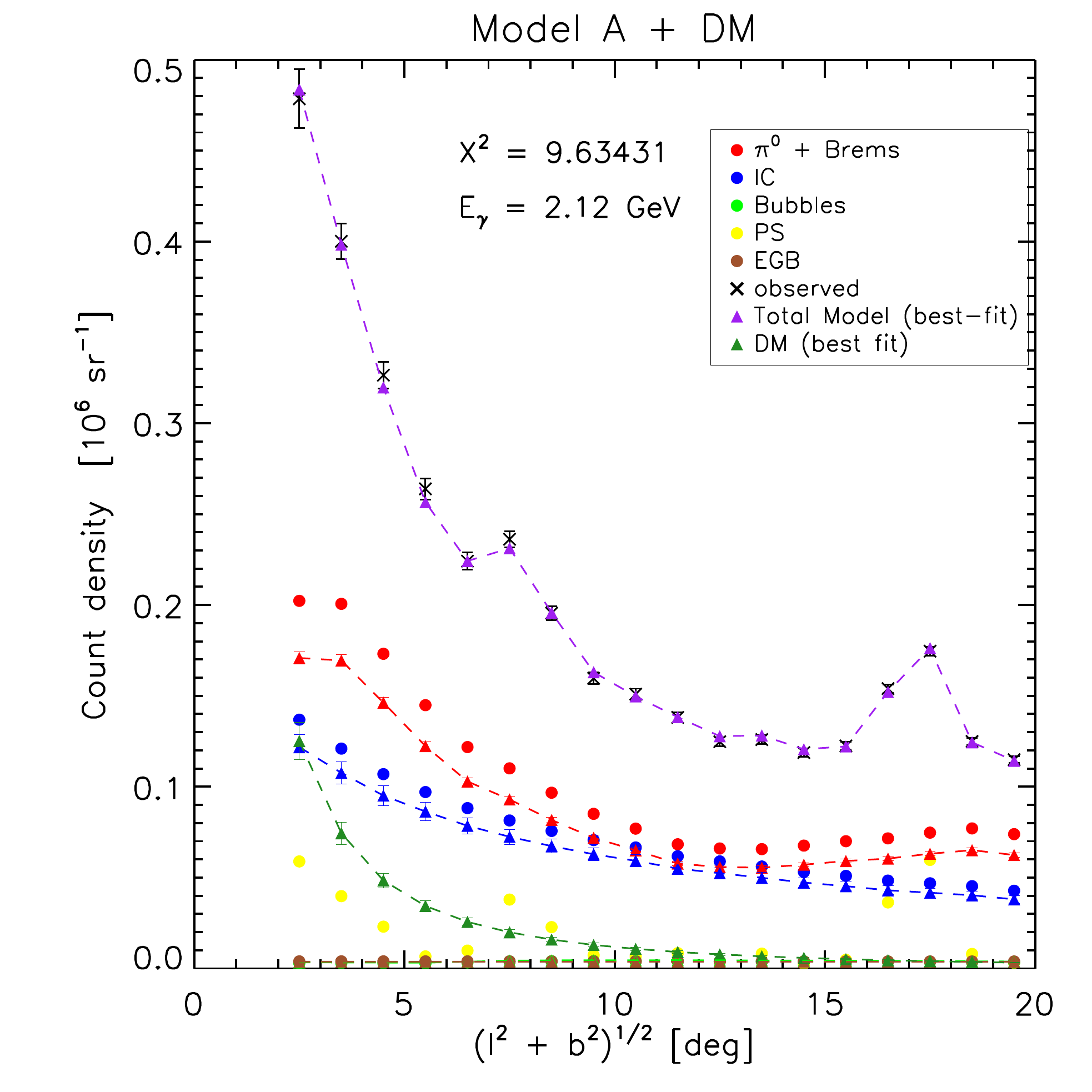}
\endminipage\hfill
\minipage{0.33\textwidth}
  \includegraphics[width=1.1\linewidth]{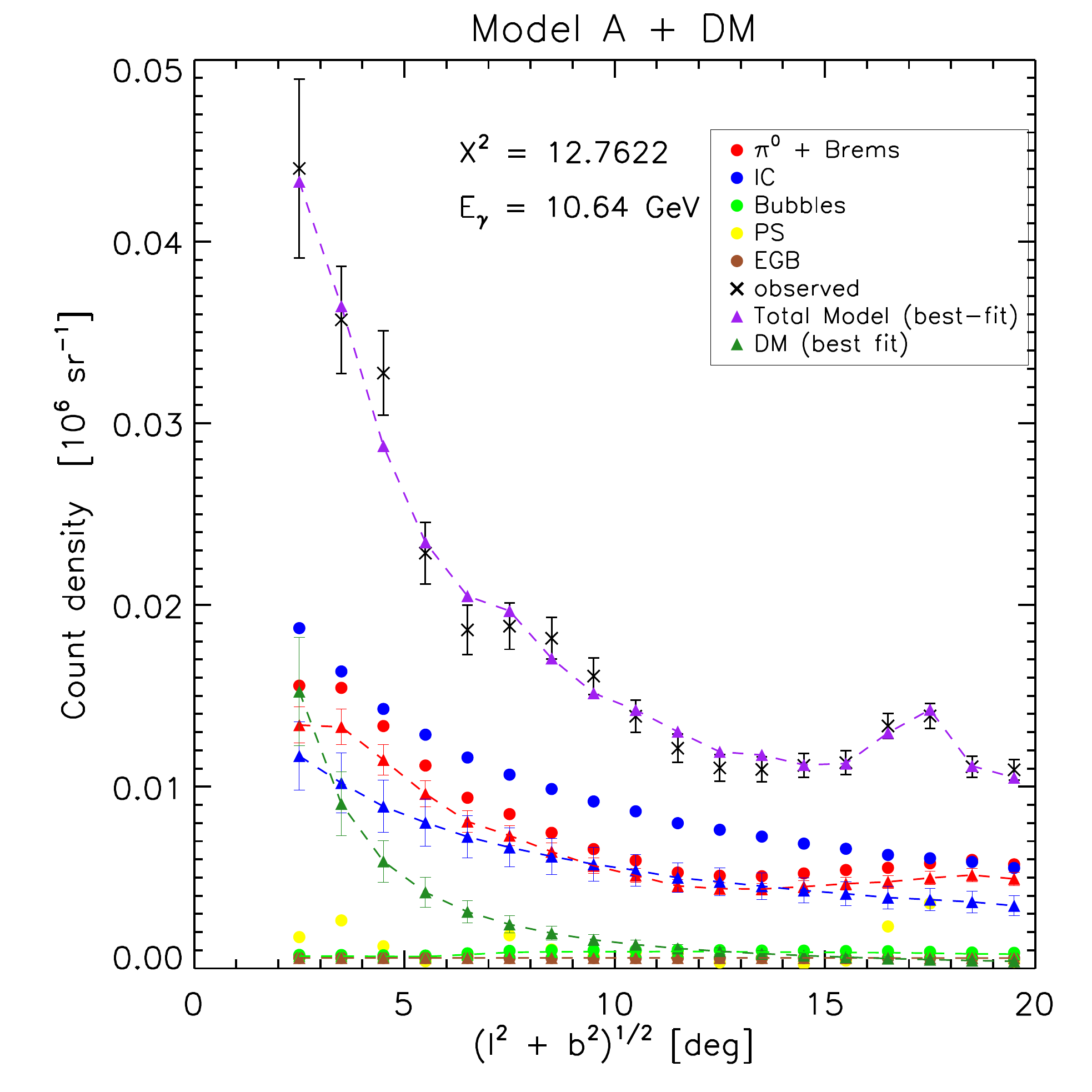}
\endminipage\\
 \caption{\small \textit{The same as fig. \ref{fig:lat_profiles} for the radial profiles.}}
 \label{fig:rad_profiles}
\end{figure}


As anticipated before, let us start from the simplest realization of our idea.

We choose $\sigma = 300$ pc as a sample value for the spike width, in agreement with astrophysical inputs. The choice of $\mathcal{N}$ is based on two objective requirements:\footnote{We refer the reader to section~\ref{sec:ParameterScanSigma} for a wider discussion about the impact of different values of $\sigma$ and $\mathcal{N}$.} The value cannot be too large, since the SFR in the GC cannot exceed few percent of the total rate in the Galaxy; moreover, we verify {\it a posteriori} that the spike emission absorbs the majority of the GC excess (to put it another way, it means that if we include in the fit of the reference model the DM template as well, the latter should be used only marginally). 

Our analysis is based on the comparison between the reference model defined before, and the case in which we add to Model A a DM template, modeled according to a modified NFW distribution with $\gamma = 1.26$ (see Sect.~\ref{sec:method} for details). The inclusion of a DM template in the fitting procedure in addition to standard astrophysical background models -- like the Model A considered in this analysis -- provided the most striking evidence supporting the DM interpretation of the GC excess. 
In~\cite{Daylan:2014rsa}, the addition of the DM template dramatically improved the fit up to an overwhelming high level of statistical significance. 
For this reason, it is mandatory to compare the performances of the spike model with those of the DM template, and we organize our analysis in three subsequent steps. 
First, in section~\ref{sec:EnergySpectrum}, we present the result of the template fit in terms of energy spectra. 
Second, in section~\ref{sec:TS}, we provide a more quantitative discussion based on the likelihood analysis. 
Finally, in section~\ref{sec:Profiles}, we offer a more direct comparison with data based on the analysis of $\gamma$-ray profiles along three complementary directions.


\subsection{Energy spectrum}\label{sec:EnergySpectrum}

As far as the energy spectra are concerned, in fig.~\ref{fig:spectrum} we present the
results of our template-fitting analysis. 

In the left panel, we show the best fit obtained for the Model A including the DM template. As described in the caption of fig. \ref{fig:spectrum}, we show for all components both pre- and post-template fitting values.
As expected, the fit heavily uses the  DM template, and the corresponding energy spectrum (magenta diamonds in fig.~\ref{fig:spectrum}) 
clearly shows the familiar bump-shaped form peaked at $E_{\gamma} = 1$\,-\,$3$ GeV, in good 
agreement with that predicted from a WIMP with $\sim 45$ GeV mass annihilating, among other possibilities, into $b\bar{b}$.
For illustrative purposes, it is instructive to look at the residual map -- i.e. the difference in counts between data and model evaluated at the best-fit point -- obtained without the inclusion of the DM template. We show this residual map in the left panel of fig.~\ref{fig:residuals} for the energy window $E_{\gamma} = 1$\,-\,$10$ GeV, where
the presence of an excess distributed around the GC stands out in full glory. This excess is fully absorbed by the DM template in the fit, as shown in the central panel of fig.~\ref{fig:residuals} where we present the residuals after template fitting including the DM template.

Let us now turn the attention to our reference model.

In the right panel of fig.~\ref{fig:residuals} we show the pattern of residuals replacing the DM template with the spike ($\mathcal{N} = 2.2$\%). The resulting pattern is analogous to the DM case.

In the right panel of fig.~\ref{fig:spectrum}, instead, we consider the possibility of having at the same time the spike and the DM component.
The contribution of the DM template is significantly reduced, is consistent with zero (within the error bars obtained from the templated fitting) in  a large energy range, and -- most importantly -- gives rise to a featureless spectrum.



In the next section we will see that precisely in this energy range, our model (without DM) gives a description of the $\gamma$-ray emission quantitatively similar to that of the Model A+DM. In other words, most of the GC excess is absorbed by the spike, and the presence of an extra DM template only gives a slight improvement of the fit.

\subsection{Test statistic}\label{sec:TS}

In order to scrutiny the performance of our model, let us now consider the upper left panel of fig.~\ref{fig:likelihood} where we compare, for each energy bin, 
the Test Statistic (TS) of the models we are studying. 
The TS is defined by ${\rm TS} = -2\Delta\log\mathcal{L}$, and is computed adopting the Likelihood $\mathcal{L}$ defined in Eq. \ref{eq:Likelihood}. 

First, we show the improvement in the TS obtained adding the DM template to the Model A. 
More precisely, the yellow dashed line with open circles in  fig.~\ref{fig:likelihood} 
represents the square root of the difference $-2\log\mathcal{L}_{\rm Model\,A}+2\log\mathcal{L}_{\rm Model\,A + DM}$ (hereafter, $\Delta$TS).
In terms of energy spectra, the case Model A + DM corresponds to the one shown 
in the left panel of fig.~\ref{fig:spectrum}. The plot clearly shows the improvement in the fit due to the presence of the DM template. 
If taken at face value, it corresponds to a statistical preference of about $15\,\sigma$ at the position of the peak. 
Of course we remark that this value should be taken {\it cum grano salis}, 
since -- in addition to the extremely small statistical errors -- data are plagued by unavoidable systematic errors not included in the likelihood fit.
However, it is indisputable that including the DM template greatly improves the fit; therefore, it is crucial to compare this result with the performance of our spike.

The green solid line with filled circles represents the 
improvement in the fit (quantified by $\Delta$TS) obtained considering our reference spike model with respect to the Model A. 
The plot highlights that our scenario -- without any DM contribution -- performs better than the starting Model A, 
and gives a result comparable (even at the level of statistical preference) with the DM case. 
For completeness,  we also show the TS for the combination of the spike and the DM template. 
The green solid line with open circles represents the corresponding $\Delta$TS, 
so that negative values indicate a statistical preference for the addition of the DM template. 
In terms of energy spectra, this situation 
corresponds to the right panel of fig.~\ref{fig:spectrum}. 
The TS plot shows that the addition of the DM template slightly improves the fit in the energy window $E_{\gamma} = 1$\,-\,$10$ GeV; 
however -- as already noticed discussing the energy spectrum in fig.~\ref{fig:spectrum} and the residual map in the right panel of fig.~\ref{fig:residuals} -- 
this improvement cannot be interpreted as the evidence of a DM contribution since the 
majority of the excess was absorbed by the presence of the spike.

Let us now pause a moment to summarize what we found. 

Looking at spectra and $\Delta$TS, we clarified that the presence of the spike -- even in the simple realization discussed here -- provides a viable astrophysical alternative to the DM interpretation of the GC excess. 

Nevertheless, both the energy spectra in fig.~\ref{fig:spectrum} and the TS in the upper left panel of fig.~\ref{fig:likelihood}
miss a transparent comparison with the $\gamma$-ray data throughout the ROI. For this reason, we decide to get
more deeply into details by analyzing latitude, longitude and radial profiles.

\subsection{Longitude, latitude, and radial profiles}\label{sec:Profiles}

In order to compare models and data, we analyze the $\gamma$-ray profiles investigating three complementary directions. 
In fig.~\ref{fig:lProfiles} we show the control regions along the Galactic latitude (left panel, red), longitude (central panel, blue) and the projected radial distance $\sqrt{l^2 + b^2}$ (right panel, green) used in our analysis.
We compute the  latitude (longitude) profiles averaging on $|l| \leqslant 5^{\circ}$ ($2^{\circ} \leqslant |b| \leqslant 5^{\circ}$), and we use $1^{\circ}$ bins. 
The Galactic disk is always masked for $|b| \leqslant 2^{\circ}$. 
We show our results for the longitudinal direction in the upper-right panel of fig.~\ref{fig:likelihood}.
Latitude and radial profiles are shown, respectively, in the lower-left and lower-right panel of fig.~\ref{fig:likelihood}.
We will come back to comment these plots in a second; to start, let us clarify some technical details.
Our procedure goes as follows: First, we compute -- for each direction, and for each energy bin --
the observed profiles of $\gamma$-ray data.
Second, using these data and the corresponding statistical errors, we compute the value of the $\chi^2$ function
for each one of the analyzed models, using the best-fit coefficients obtained from the template fitting procedure as explained in the previous section.
We repeat the analysis for the Model A, the Model A + DM and our reference model.
In figs.~\ref{fig:lat_profiles}\,-\,\ref{fig:rad_profiles} we show, for three representative energy  bins, 
the latitude, longitude and radial profiles for the reference model (top row) and the Model A + DM (bottom row).
In the three aforementioned panels of fig~\ref{fig:likelihood}, on the contrary, we superimpose the differences $\Delta\chi^2_{\rm DM} \equiv \chi^2_{\rm Model\,A}- \chi^2_{\rm Model\,A + DM}$ (empty circles)
and $\Delta\chi^2_{\rm Spike} \equiv  \chi^2_{\rm Model\,A}- \chi^2_{\rm Spike}$ (filled circles). For each line, positive values indicate 
an improvement in the fit -- compared to Model A -- obtained including, respectively, the DM template and the spike. 
In each  panel, the relative position of the two lines, $\Delta\chi^2_{\rm DM}$ versus $\Delta\chi^2_{\rm Spike}$,  indicate which one of the two ingredients, DM or spike, exhibits a better fit.
Let us now get to the heart of the matter. 
On a general ground, we will explore the three profiles separating two different energy ranges. On the one hand, 
the energy region $E_{\gamma} = 1$\,-\,$10$ GeV (mid-energy region hereafter), where the DM template gives the most pronounced contribution; on the other one, the low-energy tail of the spectrum, $E_{\gamma} = 0.3$\,-\,$1$ GeV.

\begin{itemize}

\item[$\circ$] {\it Latitude profiles.} 
These profiles show a remarkable connection with the morphology of the excess. The GC excess extends far away from the GC itself, out to at least $10^{\circ}$ in latitude \cite{Daylan:2014rsa,WeiEAlfredo}.
This is clear from the latitude profile shown in the central and right panels of the bottom row in fig.~\ref{fig:lat_profiles} where, at the reference energies $E_{\gamma} = 2.12$ GeV and 
$E_{\gamma} = 10.64$ GeV, 
the DM template gives a sizable contribution to the total count density up to $|b| \sim 10^{\circ}$.
From the central and right panels of the top row in the same figure, we see that the spike gives a comparably good fit, since the role of the DM contribution 
is entirely played by the IC emission of the spike.
This evidence is confirmed in the lower-left panel of fig.~\ref{fig:likelihood}
where it is clear that in the mid-energy region the spike provides a fit of the latitude profiles remarkably closed to one obtained by using the DM template.
As far as the low-energy region is concerned, in the lower-left panel of fig.~\ref{fig:likelihood} we observe 
that, in the first two energy bins, the DM template produces a slightly better fit if compared with the spike.
Let us now try to interpret this discrepancy.
First of all, as already pointed out previously, we remind that the purpose of these plots is to offer a
qualitative description of the comparison between models and data
 rather than derive strong statistical statements. 
 Bearing this point in mind, we turn the attention on the left panels in the bottom and top rows in  fig.~\ref{fig:lat_profiles}, where we zoom the latitude profiles in
 the second energy bin with central energy  $E_{\gamma} = 0.424$ GeV.
 The origin of the discrepancy becomes clear: while the DM contribution is suppressed at low energy (see also fig.~\ref{fig:spectrum}, left panel), the steady-state emission of the spike, being correlated to the overall IC component, cannot show this low-energy dependence (at least in the simplified model studied in this paper), and therefore leads to a slight overshooting in the region $3\lesssim |b| \lesssim 5$. 
 
 Having said that, it is undeniable that this apparently large difference ($\chi^2_{\rm Spike} - \chi^2_{\rm Model\,A + DM} \sim 50$)
 is heavily affected by the fact that we included in the computation of the $\chi^2$s only statistical errors. Indeed, the systematic errors are estimated for the integrated spectra only ($\simeq 5$\% at $E_{\gamma} = 562$ MeV~\cite{galprop2012}); a better understanding of this uncertainty will presumably change the rules of the comparison, washing out, at least partially, the discrepancy.
 
We therefore conclude that the spike provides a fit of the latitude profiles comparable in quality with the one obtained using the DM template.
  
\item[$\circ$] {\it Longitude profiles.} 
Longitude profiles are the most delicate, since they may suffer from spurious emission leaked from the masked Galactic disk. 
As before, we start our discussion from the mid-energy region, and we focus our attention on the central and right panels of the bottom row in fig.~\ref{fig:long_profiles}.
In these energy bins the bump-shaped form of the GC excess distinctly stands out.
Even in this case, we notice that the spike provides an excellent fit to data, comparable in quality to the DM template.
This is confirmed by the upper-left panel of fig.~\ref{fig:likelihood}, where 
we show the comparison between $\Delta\chi^2_{\rm DM}$ and $\Delta\chi^2_{\rm Spike}$: no significant discrepancy is reported, and in the mid-energy region  the spike 
perfectly substitutes the DM template. 

In the low-energy region the spike improves the fit 
if compared with the Model A but, as clear from the upper-left panel of fig.~\ref{fig:likelihood},
we find in the first three bins
a preference (up to $\chi^2_{\rm Spike} - \chi^2_{\rm Model\,A + DM} \sim 100$) for the DM template.
In the left panels of the bottom and top rows in  fig.~\ref{fig:long_profiles} we show the longitude profiles in
 the second energy bin with central energy  $E_{\gamma} = 0.424$ GeV.
 Clearly, the presence of the spike seems to be disfavored since it produces a visible overshoot of low-energy longitude profiles in $|l| \lesssim 4^{\circ}$.
As already stated discussing the latitude profiles, the estimate of systematic error 
is important to assess the significance of this discrepancy. Without them, no conclusive statement can be made.
Most importantly, we remark that -- from the point of view of transport properties 
-- the GC is far from being understood; non-standard properties of diffusion, 
very much likely in such a complex environment, may completely alter the simplified picture adopted so far in the description of the GC excess.
{Moreover, as mentioned in sec.~\ref{sec:method}, we point out that our results regarding the low-energy tail of the spectrum (below 1 GeV) are sensitive to the smoothing algorithm. In particular, if we choose the alternative strategy of smoothing only the model templates applying a convolution with the Fermi-LAT PSF, and leaving the point source, exposure, and count maps untouched, we find a worsening of the fit for $E < 1$ GeV.
However, due to all the caveats discussed above, that energy region is not crucial for our analysis since it is affected by major uncertaintes especially on the modelling side. Therefore, all the main results we present here are not dependent on these details.}


To conclude, we find that the spike provides a fit of longitude profiles comparable in quality with the one obtained using the DM template in the mid-energy region, where the evidence of the GC excess was considered stronger.

\item[$\circ$] {\it Radial profiles.}  Finally, we discuss the radial profiles. From the right panel in fig.~\ref{fig:lProfiles} we first notice that the adopted binning spans a wider region of the ROI compared to the previously discussed profiles.  
Indeed, the  comparison between  $\Delta\chi^2_{\rm DM}$ and $\Delta\chi^2_{\rm Spike}$  
in the lower-right panel of fig.~\ref{fig:likelihood} closely resembles the TS discussed in section~\ref{sec:TS}. 
The spike provides a very good fit, comparable in quality with the DM template both in mid- and low-energy regions.
The only exception is represented by the very first energy bin, but the same arguments outlined for latitude profiles hold true.
The radial profiles in the top and bottom rows in fig.~\ref{fig:rad_profiles} 
show in detail -- considering the usual benchmark bins at $E_{\gamma} = 0.424,\,2.12\,,10.64$ GeV -- 
the performance of the spike with respect to the DM template. It is clear as the modified IC emission, altered by the presence of the spike, mimics the DM   
contribution up to $\sqrt{l^2 + b^2} \sim 10^{\circ}$.

\end{itemize}

To sum up,
 the analysis of latitude, longitude and radial profiles 
enforces what already found in section~\ref{sec:TS}. 
 We confirm that the presence of the spike 
in our reference model 
depicts a viable astrophysical scenario potentially able to fully explain the GC excess.  




\section{Discussion}
\label{sec:ParameterScanSigma}



\begin{figure}[!htb!]
\minipage{0.48\textwidth}
  \includegraphics[width=1.\linewidth]{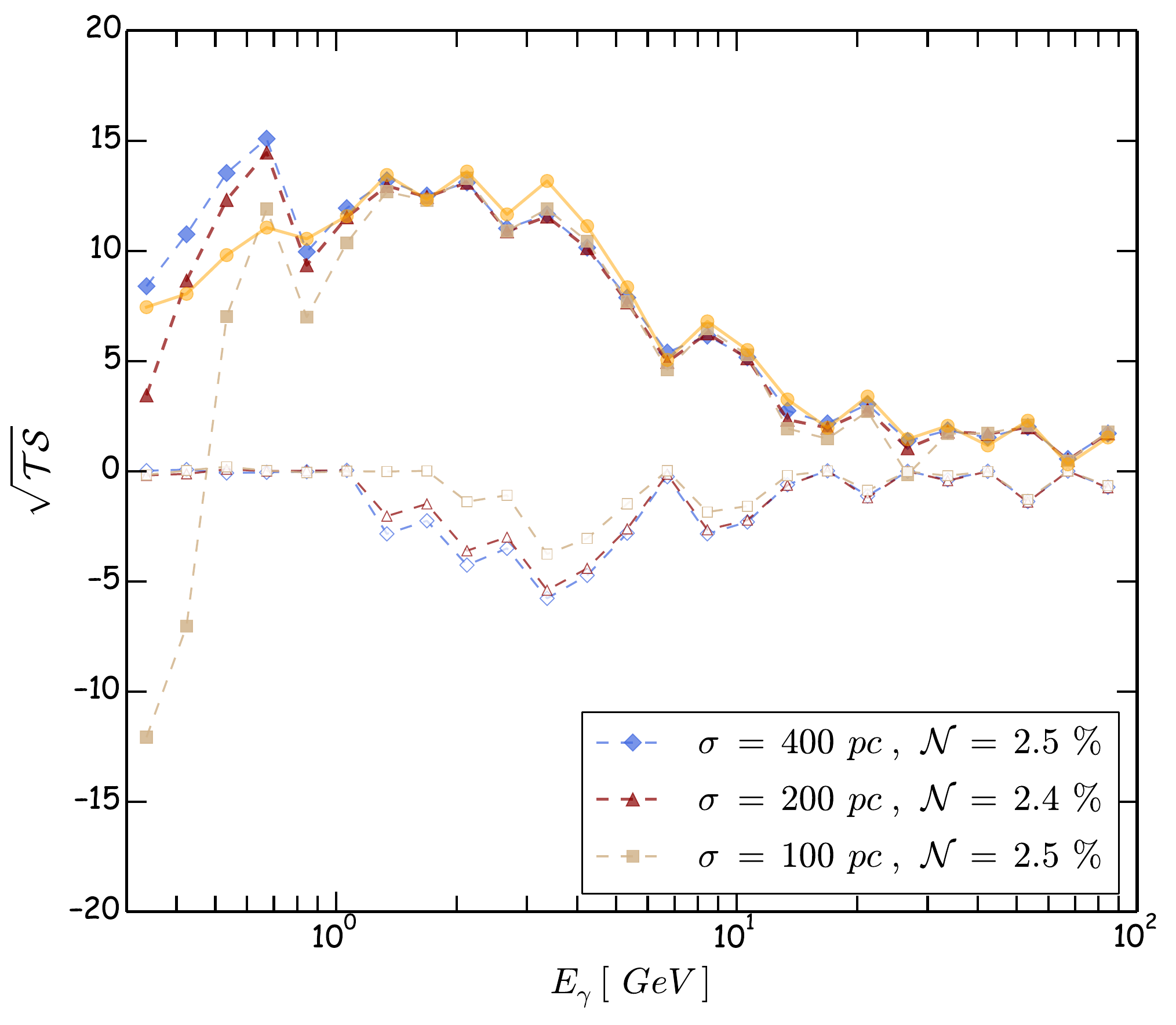}
  \caption{TS plot for $\sigma = 100 \div 400$ pc. }\label{fig:TS_scan_sigma}
\endminipage\hfill
\minipage{0.48\textwidth}
  \includegraphics[width=1.\linewidth]{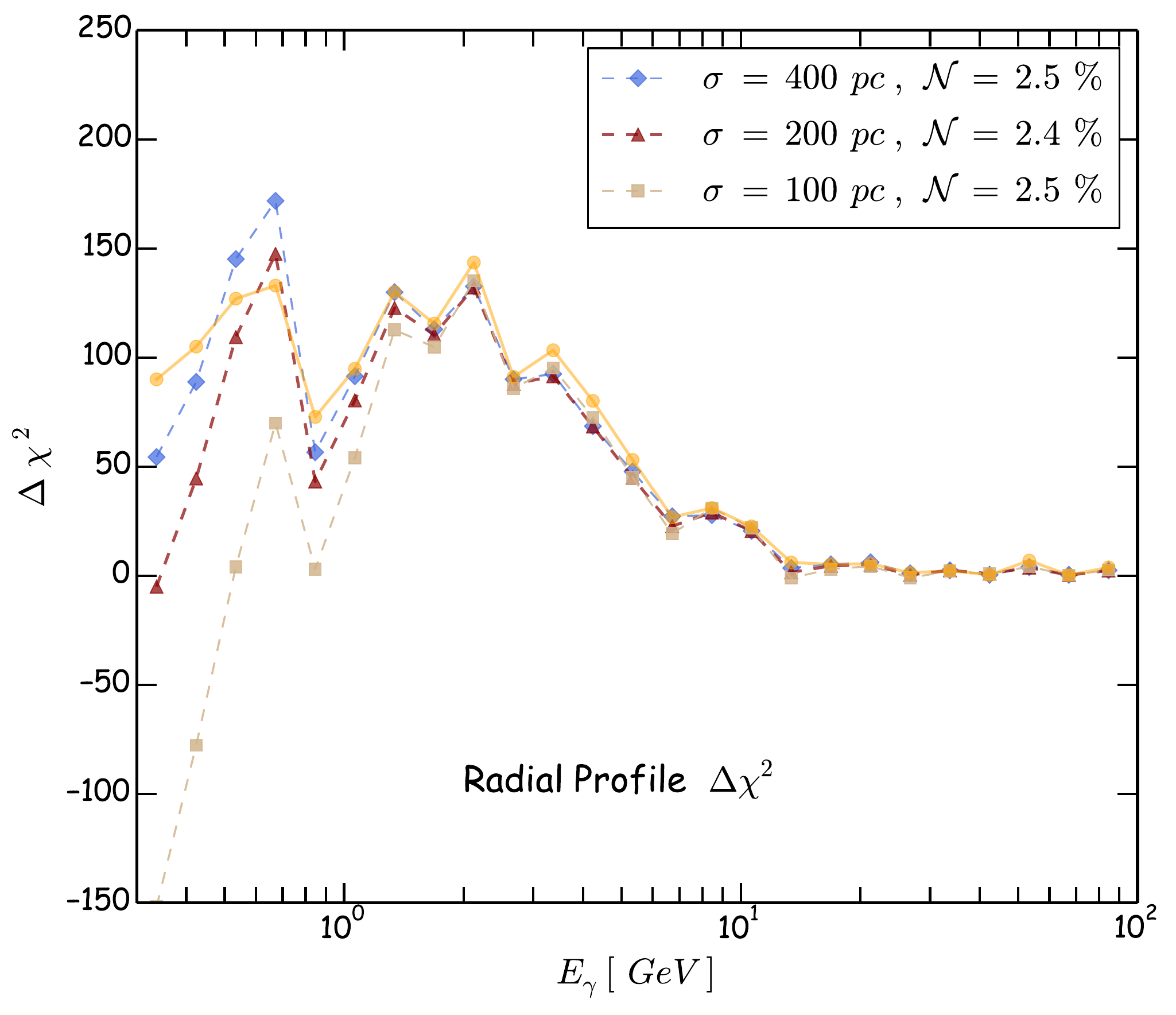}
  \caption{Radial $\chi^2$ for $\sigma = 100 \div 400$ pc. }\label{fig:DeltaChi2_rad_sigma}
\endminipage\\
\label{fig:spectra}
\end{figure}

Although we presented in detail a viable working scenario, it is worth mentioning that several variations are possible.

\begin{itemize}

\item
First of all, we show that the nice agreement with the data obtained for the reference case, can be extended to different values of $\sigma$ (see fig.~\ref{fig:TS_scan_sigma} and fig.~\ref{fig:DeltaChi2_rad_sigma}).  
 
In particular we verified that, going down to $\sigma = 200$ pc and up to $\sigma = 400$ pc (with a proper rescaling of $Q_0$ in Eq. \ref{eq:spike}), all the results of the previous section are almost unchanged. 
Interestingly, in all cases the optimal value of $\mathcal{N}$ does not change much. 

We remark that spikes more extended than those considered here are in tension with astrophysical observations, since -- as discussed in the Introduction -- the evidence for a significant star-forming activity is confined in the inner part of the Galactic bulge, and the SFR at $R\simeq0.5$ kpc is expected to steeply decrease.

On the other hand, for small values of $\sigma$, below $200$ pc, the fits to the low- up to intermediate-energy data significantly worsen, indicating an unsatisfactory description of the $\gamma$-ray emission: A narrow spike, indeed, exacerbates the problem highlighted in the previous section with longitude profiles.


The origin of the problem is the shape of the IC emission: once we fix the normalization of the spike in order to match the $\gamma$-ray emission at energies above $\simeq$ 1 GeV, the IC emission for a narrow spike is significantly steeper than larger values of $\sigma,$ as shown in fig.~\ref{fig:IC_profile}.
Such morphology is disfavored by the low energy data.


\item

In order to provide a more realistic description of the inner Galaxy, one should take into account complex interplays in the GC environment impacting on CR tranport properties.  

In particular, we know that the regular Galactic magnetic field can be decomposed into several components: besides a well-known pattern lying on the Galactic plane, in a recent detailed analysis \cite{farrar} the authors pointed out a X-shaped component extending from the GC up to a Galactocentric radius $R \simeq 4 \div 5$ kpc and to $z \simeq 2 \div 3$ kpc. 
Since CR diffusion is expected to be anisotropic, and since -- according to the Quasi-Linear Theory (QLT) valid in the low-turbulence limit -- the parallel diffusion should dominate over the perpendicular one, we expect an efficient CR escape due to parallel diffusion in the vertical direction. 

We checked several values of the ratio between the perpendicular and parallel diffusion components of the diffusion tensor ($D_{zz}/D_{RR}$) for each value of $\sigma$, but we did not find any relevant and solid improvement in the quality of the fit valid in the whole energy range; 
We postpone to a more detailed forthcoming work a careful study regarding the impact of this ingredient.


\item 


Although we chose Model A as a reference case, we verified that our results are not strongly dependent on the CR transport model.

The same trends we discussed in detail in the previous section emerge if other CR propagation models are adopted.
In particular, we considered the conventional KRA model described in \cite{Evoli:2011id}, and Model F in \cite{Calore}. In both cases, we find that the spike plays a crucial role when the template-fitting algorithm is applied, and all the results presented in the previous section still hold. 

\end{itemize}

\section{Conclusions}

This paper investigates the possibility of addressing the $\gamma$-ray Galactic center excess, as pointed out in the literature and possibly associated to a DM annihilation component, in terms of ordinary SNR sources and standard steady state CR diffusion. A new reference model is introduced including a population of SNRs in the central region of the Galaxy; while such term was previously overlooked in CR propagation modelling, it is strongly motivated based on the observed level of star formation rate and supernova explosions in the central region of the Galaxy. The main impact of this extra source in the Galactic center analysis is to enhance the IC emissivity, introducing a morphological imprint analogous -- but different when coming to details -- to DM pair annihilations in a spherical NFW-like density profile. 



Differently from previous proposals for alternative scenarios, we introduced the new ingredient in the context of the template-fitting algorithm, treating  the new contribution as correlated to the conventional IC emission. We analyzed in detail the overall goodness of the fit of our framework, and produced a more direct comparison against data examining profiles in different directions. 
{Remarkably, the test statistic of the fit related to our scenario turns out to be as good as the Dark Matter one in the ROI here considered.}


Our result is solid against different point-source masking technique, {standard} choices of the CR transport properties, and holds for a central source width ranging from $\simeq 200$ to $\simeq 400$ pc.

The additional source proposed here provides extra $\gamma$-ray emissivity also at energies larger than the GeV range analyzed in this work, although with a progressively smaller angular extent given that, due to the increase in the energy loss efficiency, electron diffusion is reduced on a shorter scale. This prediction can be tested in the near future, in the perspective of comparing low- and high-energy measurements by experiments such as ASTROGAM~\cite{astrogam} and GAMMA-400~\cite{Topchiev:2015zha}.



Although the template-fitting method is very powerful in addressing morphological agreements or flaws of a given theoretical model, there are some drawbacks as well in this approach: Since the free-floating of the templates is performed independently on each energy bin, the $\gamma$-ray spectra corresponding to the different components can be seriously altered, resulting sometimes in little control on the way the physical properties of the CR transport model are modified. Motivated by these considerations, as a further development of this work, we plan to exploit an alternative approach in which the original features of a physical propagation model -- supposed to provide also a good fit of the local observables -- are kept under control by the data fitting procedure.


\section*{Acknowledgements}

We thank Francesca Calore, Ilias Cholis, Marco Cirelli, Carmelo Evoli, Christoph Weniger for many useful discussions.

D.G. acknowledges the SFB 676 research fellowship from the University of Hamburg as well as the hospitality of DESY. 

M.T. is supported  by the European Research Council ({\sc Erc}) under the EU Seventh Framework Programme (FP7/2007-2013) / {\sc Erc} Starting Grant (agreement n. 278234 - `{\sc NewDark}' project). M.T. acknowledge the hospitality of the Institut d'Astrophysique de Paris, where part of this work was done.

The work of A.U. is supported by the ERC Advanced Grant n$^{\circ}$ $267985$, ``Electroweak Symmetry Breaking, Flavour and Dark Matter: One Solution for Three Mysteries'' (DaMeSyFla).

P.U. and M.V. acknowledge partial support from the European Union FP7 ITN INVISIBLES (Marie Curie Actions, PITN-GA-2011-289442), and partial support by the research grant ``Theoretical Astroparticle
Physics'' number 2012CPPYP7 under the program PRIN 2012 funded by the Ministero dell'Istruzione, Universit\`a e della Ricerca (MIUR).

In this work the authors have used the publicly available {\tt Fermi-LAT} data and Fermi Tools archived at \url{http://fermi.gsfc.nasa.gov/ssc/}.


\end{document}